\documentclass[11pt]{article}
\usepackage{jheppub}
\pdfoutput=1
\usepackage{graphicx}
\usepackage{fixltx2e}
\usepackage{slashed}
\usepackage[dvipsnames,table]{xcolor}
\usepackage{hyperref} 
\usepackage{xspace}
\usepackage[tight]{subfigure}
\usepackage{amsmath}
\usepackage{amssymb}
\usepackage{amsfonts}
\usepackage{mathrsfs}
\usepackage{comment}
\usepackage{afterpage}
\usepackage{verbatim}
\usepackage{booktabs}
\usepackage{array}
\usepackage[title,titletoc]{appendix}
\usepackage{tabularx}
\newcolumntype{C}[1]{>{\centering\arraybackslash}p{#1}}

\def\refse#1{\mbox{Section~\ref{#1}}}

\def\citere#1{\mbox{Ref.~\cite{#1}}}

\newcommand{\newc}{\newcommand}
\newc{\beq}{\begin{equation}}
\newc{\eeq}{\end{equation}}
\newc{\bit}{\begin{itemize}}
\newc{\eit}{\end{itemize}}
\newc{\ben}{\begin{enumerate}}
\newc{\een}{\end{enumerate}}
\newc{\bce}{\begin{center}}
\newc{\ece}{\end{center}}
\newc{\bfi}{\begin{figure}}
\newc{\efi}{\end{figure}}

\newcommand{\ri}{\mathrm i}

\newcommand{\rd}{\mathrm d}

\newcommand{\rT}{{\mathrm{T}}}

\newcommand{\rL}{{\mathrm{L}}}

\newcommand{\eg}{\emph{e.g.}\ }

\renewcommand{\Re}{\mathop{\mathrm{Re}}\nolimits}

\newcommand{\GeV}{\ensuremath{\,\text{GeV}}\xspace}

\newcommand{\PH}{\ensuremath{\text{H}}\xspace}

\newcommand{\Pp}{\ensuremath{\text{p}}}
\newcommand{\Pe}{\ensuremath{\text{e}}\xspace}

\newcommand{\Pt}{\ensuremath{\text{t}}\xspace}

\newcommand{\PW}{\ensuremath{\text{W}}\xspace}
\newcommand{\PZ}{\ensuremath{\text{Z}}\xspace}

\newcommand{\Mt}{\ensuremath{m_\Pt}\xspace}
\newcommand{\MH}{\ensuremath{M_\PH}\xspace}
\newcommand{\MWOS}{\ensuremath{M_\PW^\text{OS}}\xspace}

\newcommand{\MZOS}{\ensuremath{M_\PZ^\text{OS}}\xspace}
\newcommand{\MZ}{\ensuremath{M_\PZ}\xspace}

\newcommand{\GZ}{\ensuremath{\Gamma_\PZ}\xspace}

\newcommand{\GF}{\ensuremath{G_\mu}}

\newcommand{\recola}{{\sc Recola}\xspace}

\newcommand{\mocanlo}{{\sc MoCaNLO}\xspace}

\newcommand{\collier}{{\sc Collier}\xspace}

\newcolumntype{.}{D{.}{.}{-1}}
\newcolumntype{d}[1]{D{.}{.}{#1}}
\colorlet{tableoverheadcolor}{gray!37.5}
\colorlet{tableheadcolor}{gray!25}
\colorlet{tablerowcolor}{gray!12.5}

\def\draftdate{\relax}
\def\mda{\relax}
\def\mua{\relax}
\def\mla{\relax}
\def\draft{
\def\thtystars{******************************}
\def\sixtystars{\thtystars\thtystars}
\typeout{}
\typeout{\sixtystars**}
\typeout{* Draft mode!
         For final version remove \protect\draft\space in source file *}
\typeout{\sixtystars**}
\typeout{}
\def\draftdate{\today}
\def\mua{\marginpar[\boldmath\hfil$\uparrow$]%
                   {\boldmath$\uparrow$\hfil}\color{black}%
                    \typeout{marginpar: $\uparrow$}\ignorespaces}
\def\mda{\color{red}\marginpar[\boldmath\hfil$\downarrow$]%
                   {\boldmath$\downarrow$\hfil}%
                    \typeout{marginpar: $\downarrow$}\ignorespaces}
\def\mla{\marginpar[\boldmath\hfil$\rightarrow$]%
                   {\boldmath$\leftarrow $\hfil}%
                    \typeout{marginpar: $\leftrightarrow$}\ignorespaces}
\def\Mua{\marginpar[\boldmath\hfil$\Uparrow$]%
                   {\boldmath$\Uparrow$\hfil}\color{black}%
                    \typeout{marginpar: $\uparrow$}\ignorespaces}
\def\Mda{\color{red}\marginpar[\boldmath\hfil$\Downarrow$]%
                   {\boldmath$\Downarrow$\hfil}%
                    \typeout{marginpar: $\downarrow$}\ignorespaces}
\def\Mla{\marginpar[\boldmath\hfil\textcolor{red}{$\Rightarrow$}]%
                   {\boldmath\textcolor{red}{$\Leftarrow $}\hfil}%
                    \typeout{marginpar: $\leftrightarrow$}\ignorespaces}
\overfullrule 5pt
\oddsidemargin 15mm
\marginparwidth 29mm
}

\newcommand{\Deltaf}{F_{\pm}}
\newcommand{\mc}{\mathcal}

\newcommand{\pt}[1]{p_{\rT}({#1})}

\newcommand{\nnb}{\nonumber}

\newcommand{\sweff}{\sin^2\!\theta_{\textrm{w}}^{\,\textrm{eff}}}

\begin{document}


\title{Z-boson quantum tomography at next-to-leading order}

   \author{Morgan Del Gratta$\,^{a,b}$,}\emailAdd{morgan.delgratta@unimi.it}
   \author{Federica Fabbri$\,^{c,d}$,}\emailAdd{federica.fabbri@cern.ch}
   \author{Michele Grossi$\,^{e}$,}\emailAdd{michele.grossi@cern.ch}
   \author{Fabio Maltoni$\,^{c,d,e,f}$,}\emailAdd{fabio.maltoni@unibo.it}
   \author{Davide Pagani$\,^{d}$,}\emailAdd{davide.pagani@bo.infn.it}  
   \author{Giovanni Pelliccioli$\,^{g,h,1}$\note{Corresponding author}, }\emailAdd{giovanni.pelliccioli@unimib.it}
   \author{Alessandro Vicini$\,^{a,b}$}\emailAdd{alessandro.vicini@mi.infn.it}

    \affiliation{\vspace*{0.3cm}$\,^{a}$University of Milano, Department of Physics, 20133 Milano, Italy}
    \affiliation{$\,^{b}$INFN, Sezione di Milano, 20133 Milano, Italy}
   \affiliation{$\,^{c}$University of Bologna, Department of Physics and Astronomy, 40126 Bologna, Italy}
   \affiliation{$\,^{d}$INFN, Sezione di Bologna, 40126 Bologna, Italy}
      \affiliation{$\,^{e}$European Organisation for Nuclear Research (CERN), 1211 Geneva, Switzerland}
      \affiliation{$\,^{f}$Centre for Cosmology, Particle Physics and Phenomenology (CP3), Universit\'e Catholique de
Louvain, B-1348 Louvain-la-Neuve, Belgium}
      \affiliation{$\,^{g}$University of Milano--Bicocca, Department of Physics, 20126 Milano, Italy}
   \affiliation{$\,^{h}$INFN, Sezione di Milano--Bicocca, 20126 Milano, Italy}

 \abstract{
 We investigate the origin of the unusually large electroweak (EW) radiative effects observed in the extraction of the spin--density matrix and related observables at colliders, focusing on leptonic $\PZ$-boson decays. We compute the $\PZ$-boson decay spin-density matrix at next-to-leading order (NLO) and find that, while its analytic structure remains essentially unchanged with respect to leading order, the EW corrections induce a sizeable $-35\%$ shift in the spin-analysing power parameter $\eta_\ell$. This effect alone accounts for the striking size of the corrections. For boosted $\PZ$ bosons, we further show that the treatment of photon radiation in lepton-dressing algorithms significantly affects the extraction of spin-density-matrix coefficients at NLO and must be carefully controlled. To address these challenges, we propose a quantum tomography procedure that is applicable to any final state with one or more on-shell $\PZ$ bosons that is robust under higher-order corrections. We illustrate its validity and limitations in $\Pp\Pp \to \PZ\PZ \to 4\ell$ and in heavy ($\MH>2 \MZ$) Higgs boson decay $\PH\to \PZ\PZ \to 4\ell$.
}

\preprint{COMETA-2025-41}
\maketitle

\section{Introduction}\label{sec:intro}

The growing interest in accessing quantum-information inspired observables at collider energies \cite{Barr:2024djo,Afik:2025ejh}, also spurred by the recent ATLAS and CMS observation of the entanglement between the spin of top-quarks~\cite{ATLAS:2023fsd,CMS:2024pts,CMS:2024zkc},
has triggered numerous studies in other channels, amongst which a significant role is played by final states with two electroweak (EW) bosons in the final state \cite{Barr:2021zcp,Aguilar-Saavedra:2022wam,Ashby-Pickering:2022umy,Aguilar-Saavedra:2022mpg,Fabbrichesi:2023cev,Fabbrichesi:2023jep,Aoude:2023hxv,Bernal:2023ruk,Fabbri:2023ncz,Morales:2023gow,Bernal:2024xhm,Grossi:2024jae,Sullivan:2024wzl,Wu:2024ovc,Grabarczyk:2024wnk,Aguilar-Saavedra:2024jkj,Subba:2024aut,DelGratta:2025qyp,Ding:2025mzj,Aguilar-Saavedra:2025byk,Goncalves:2025qem,Ruzi:2025jql,Goncalves:2025xer}.
A central tool for accessing quantum-inspired observables related to the spin of particles produced at colliders is the \emph{quantum tomography} (QT) procedure. This method enables the reconstruction of the full or partial spin-density matrix, which encapsulates the polarisation and spin-correlation information of the system. For example, the case of a pair of massive electroweak bosons can be thought of as a two-qutrit system, {\it i.e.}, a bipartite quantum system composed of two three-level subsystems. In the most general case, such a system is characterised by eighty independent polarisation and spin-correlation coefficients, whose values are sensitive to the underlying production dynamics. QT provides a systematic framework to extract these coefficients by analysing the angular distributions of the bosons decay products; see \eg \citere{Barr:2024djo}.

The standard QT procedure is based on  leading order (LO) computations of scattering amplitudes. However, the data that are being collected at the LHC include also effects due to higher-order corrections, both loop and real radiation, in the QCD and EW couplings. It is therefore mandatory to study the impact that such higher order effects may have on the extracted spin density matrix.
The first studies of such corrections in the context of two-boson spin correlations at colliders, {\it i.e.} systems of two qutrits in the quantum-information wording, are very recent \cite{Grossi:2024jae,DelGratta:2025qyp,Goncalves:2025qem,Aguilar-Saavedra:2025byk,Goncalves:2025xer}.   It has been shown for Higgs-boson decays to four charged leptons \cite{Grossi:2024jae,DelGratta:2025qyp,Goncalves:2025qem} that higher-order and off-shell effects induce very large corrections, making the spin-density matrix for two-qutrit systems not well-defined and its interpretation in terms of spin entanglement and Bell-inequality violation possibly problematic. This issue arises from large corrections to the spin analysing power $\eta_\ell$, a key quantity that relates the spin of the parent particle to the directions of the decay products and enters in the  extraction of the coefficients of the spin density matrix. This problem is particularly severe in the $Z\rightarrow \ell^+ \ell^-$ process due to the small $\eta_\ell$ value in this process. 
The first insightful proposal to solve the problem has been made in \citere{Goncalves:2025qem}, where it was noted that the large part of NLO EW effects could be incorporated at LO via a change of scheme, {\it i.e.}  using $\sweff$ \cite{Kennedy:1988sn,Renard:1994ay,Ferroglia:2001cr,Ferroglia:2002rg,Chiesa:2019nqb,Amoroso:2023uux,Biekotter:2023vbh,Chiesa:2024qzd}.

In this work we further investigate and fully understand the effects of the radiative EW corrections on the QT procedure for $\PZ$ bosons decays and propose a general solution.

First, we perform the complete calculation at next-to-leading-order (NLO) of all entries of the decay spin-density matrix for a $\PZ$ boson decaying into two charged leptons. Via a numerical analysis, we prove that the LO analytic structure of the decay matrix is not changed when considering NLO EW effects and relying on dressed charged leptons, up to a sizeable modification that can be fully absorbed into the spin-analysing power. 
This allows us to fully generalise the approach to any process involving an arbitrary number of on-shell $\PZ$ bosons in the final state, such as double and triple vector boson production, associated production, $t\PZ$, $t\bar{t}\PZ$ and so on. 

Second, we show how the derived decay-only results can be consistently applied also to the case where final state $\PZ$ bosons are boosted. In such a situation, the details of the lepton-dressing algorithm affect the evaluation of the spin-analysing power and make the extraction of the spin-density matrix coefficients rather delicate at NLO. Employing a pole-approximation strategy and selecting helicity states in tree-level and one-loop amplitudes, we derive a sound prescription to perform a QT of two-qutrit systems that is correct at NLO EW and systematically improvable at higher orders.

Third, we show how in the case of the Higgs decay into two  $\PZ$ bosons, subsequently decaying into leptons, the fact that one of the two $Z$ must be off-shell affects the aforementioned prescription, which instead would still be valid for a heavier Higgs with $\MH\gg 2 \MZ$. Such academic scenario is also exploited in order to show how effects formally of higher order (NNLO EW) and corresponding to NLO EW corrections to both decays are sizeable and we support our prescription for the extraction of the coefficients of the $\PZ\PZ$ spin density matrix.

The paper is organised as follows. In Sec.~\ref{sec:QTvsHA} we formally introduce the problem and define relevant quantities at LO and NLO EW accuracy. In Sec.~\ref{sec:decay} we consider the decay of a $\PZ$ boson (in its rest frame) and show how EW radiative effects modify the diagonal and off-diagonal entries of the decay matrix. The effects of the lepton-dressing algorithm, in association with cuts at production level in LHC processes are discussed in Sec.~\ref{sec:dress}, exploiting the helicity-amplitude method. In Sec.~\ref{sec:ZZ} we discuss in detail the structure of NLO EW corrections in the production + decay process for the $\PZ\PZ$ production.  We compare results obtained via QT and helicity-amplitude methods for quantum-information observables at LO and NLO EW accuracy. We derive via this comparison the correct prescription that should be employed in the case of QT.  In Sec.~\ref{sec:HZZ} we discuss the case of the Higgs decay, for which problems persists, and we further support our prescription for the case of $\PZ\PZ$ production and a possible heavier Higgs, highlighting the relevance of higher-order corrections.  In Sec.~\ref{sec:conclu} we draw our conclusions and provide an outlook.

\section{Quantum tomography and helicity amplitudes}

\label{sec:QTvsHA}
In quantum mechanics, the spin-density matrix of a multipartite system composed of arbitrary-spin (qudit) states encodes its most general quantum spin configuration, whether pure or mixed. In particle physics, fermions and massless bosons correspond to qubits, while massive vector bosons such as the \PW and \PZ act as qutrits. It is therefore of interest to study quantum correlations among final states produced in high-energy collisions, which can be computed perturbatively from first principles using QFT and in the present case, the Standard Model (SM) Lagrangian. Because the physical massive vector bosons are unstable and decay well before their propagation can be directly measured in the detector, only the momenta of their decay products can be observed. The final spin state of each boson, and of the composite system as a whole, must therefore be inferred statistically from the distributions of decay products. This procedure  relies on several assumptions. The first is that the EW bosons are on their mass shell. This can be enforced by selecting events in which the invariant mass of the decay products matches the mass of the corresponding bosons. From a theoretical perspective, this condition allows the amplitude to be factorised into production and decay parts and ensures that a well-defined spin-density matrix can be determined to all orders in perturbation theory.
Experimentally, however, reconstructing the spin of the original vectors from the directions of their decay products introduces complications. An obvious issue is the possible presence of additional radiation in the event. First, one must identify whether this radiation originates from the production or the decay process, a task that can be performed unambiguously only if the energy of the radiation is larger than the width of the decaying particle. In addition, it requires using the invariant-mass condition, something that can be done only if the detector has excellent energy-momentum resolution. Finally, in the case of final-state radiation, its contribution must be correctly incorporated into the tomography procedure. Moreover, when comparing theory with experiment, higher-order virtual effects must also be included when relevant. A robust procedure that enables the reliable reconstruction of the overall spin-density state of the heavy system and its comparison with theoretical predictions is therefore essential.
In this section we introduce and compare two complementary approaches to extracting the spin-density matrix of a pair of EW bosons decaying into a four-lepton final state, with inclusiveness over additional radiation (in the present case, QED). Both methods—Quantum Tomography (QT) and Helicity Amplitudes (HA)—assume, explicitly or implicitly, that the process is dominated by two on-shell spin-1 bosons, yet they offer distinct perspectives.

QT defines a finite set of observables over the full phase space by projecting onto the elements of a spherical-harmonics basis of rank $\leq 2$, whose coefficients are directly related to the spin-density matrix. While this definition can always be applied to experimental data, interpreting the results requires care. Practical limitations such as incomplete phase-space coverage from fiducial or selection cuts, the presence of real radiation, and non-resonant contributions may invalidate the assumption that a few spherical harmonics suffice. In current analyses, the data are unfolded but interpreted through a leading-order (LO) calculation.

The HA method computes the cross sections for each helicity configuration of the intermediate bosons, providing direct control over the contribution of every spin state to the total cross section and hence to the spin-density matrix. Decay kinematics and additional quantum effects are incorporated separately for each helicity, yielding a more explicit link between theoretical predictions and experimental measurements.

Although the two approaches are equivalent at LO, their behaviour differs once higher-order corrections are included. A comparison at next-to-leading order (NLO) therefore quantifies the respective advantages and limitations, guiding the choice of the most accurate strategy to determine the spin-density matrix. We now start by presenting the elements and methods proper of a LO analysis and then at move to NLO.

\subsection{Leading order}

If we consider the $\PZ+X$ production process, with the $\PZ$ boson decaying into two fermions, at LO and assuming the strict narrow-width-approximation (NWA), {\it i.e.} forcing the $\PZ$ boson to be on-shell,  the angular distribution of one of the fermions can be written as
\beq\label{eq:RhoGamma}
\frac{4\pi}3\frac1{\sigma}\frac{\rd^2\sigma}{\rd\cos\theta\,\rd\phi}\,=\,
\,\sum_{\lambda,\lambda'}\rho_{\lambda,\lambda'}
\Gamma_{\lambda,\lambda'}(\theta,\phi)
\,,
\eeq
where $\sigma$ is the total cross section, $\rho$ is the spin-density matrix of the $\PZ+X$ system traced over the spin-indices of $X$, and $\Gamma$ is the spin-density decay matrix of the $\PZ$, which depends on the polar ($\theta$) and the azimuthal ($\phi$) decay angles of one of the two fermions in the corresponding $\PZ$-boson rest frame. In the following, for simplicity, we will refer to $\Gamma$ as the ``decay matrix'' and we will consider the leptonic decay $\PZ\to \ell^+ \ell^-$, where $\theta$ and $\phi$ refer to the positively charged lepton $\ell^+$ in the $\PZ$-boson rest frame. The indices $\lambda,\lambda'=0,1,-1$  label the physical polarisation states associated to the $\PZ$ boson (longitudinal, left- and right-handed).
When also the other particles produced in association with the $\PZ$ are considered and their spin-indices are not traced, Eq.~\eqref{eq:RhoGamma} can exhibit the full form of the spin-density matrix. For instance, in the case of the $\PZ\PZ$ bipartite qutrit system, with both $\PZ$ bosons decaying leptonically, 
one obtains \beq\label{eq:RhoGammaGamma}
\frac{16\pi^2}9\frac1{\sigma}\frac{\rd^4\sigma}{\rd\cos\theta_1\,\rd\phi_1\,\rd\cos\theta_2\,
\rd\phi_2}\,=\,
\,\sum_{\lambda_1,\lambda_1',\lambda_2,\lambda_2'}\rho_{\lambda_1,\lambda_1',\lambda_2,\lambda_2'}
\Gamma_{\lambda_1,\lambda_1'}(\theta_1,\phi_1)
\Gamma_{\lambda_2,\lambda_2'}(\theta_2,\phi_2)\,,
\eeq
where the subscripts 1 and 2 label the quantities associated to the two $\PZ$ bosons.\footnote{Note that in this work we consider final state fermions in the two decays distinguishable, {\it i.e.}, we choose different flavours. While this constraint can be lifted, it simplifies the discussion in the following and most of the time is not relevant. We will further comment  on this when necessary.} 

At LO and assuming NWA, the spin-density matrix can be extracted in two different manners:
\begin{itemize}
\item QT: via the angular distributions of the positively\footnote{The use of the negatively charged leptons yields equivalent results, up to an overall sign flip in the entries sensitive to $\eta_\ell$.} charged leptons it is possible to reconstruct the  spin-density matrix. In short, by projecting the angular distributions onto spherical harmonics $Y_{lm} (\theta, \phi)$, it is possible to derive the individual entries of $\rho$.
\item HA: the individual quantities $\rho_{\lambda,\lambda'}$ and $\Gamma_{\lambda,\lambda'}$ entering Eq.~\eqref{eq:RhoGamma} can be computed either separately or with contracted helicity indices, depending on the on-shell approximation used for the calculation. More generally, the same argument can be repeated for the case of the density matrix of a bipartite $\PZ\PZ$ system, as in Eq.~\eqref{eq:RhoGammaGamma}. 

\end{itemize}

It is clear that at LO, assuming NWA and imposing no cuts on the lepton momenta, the calculation of $\rho$ via QT or HA leads to exactly the same results. For a generic $\PZ+X$ system, following the notation of \citere{Grossi:2024jae}, Eq.~\eqref{eq:RhoGamma} can be written as 
\begin{align}\label{eq:Vang}
\frac1 {\sigma}\frac{\rd \sigma}{
\rd\!\cos\theta
\,\rd\phi
} & = 
\frac{1}{4\pi}+ \, \sum_{l=1}^2 \sum_{m=-l}^{l}\, \alpha_{lm} \, Y_{lm}\big (\theta, \phi\big ) \,,
\end{align} 
and for the particular case of a $\PZ\PZ$ pair,  Eq.~\eqref{eq:RhoGammaGamma} can be  written as 
\begin{align}\label{eq:VVang}
\frac1 {\sigma}\frac{\rd \sigma}{
\rd\!\cos\theta_{1}
\,\rd\phi_{1}
\,\,\rd\!\cos\theta_{2}
\,\rd \phi_{2}
} & = 
\frac{1}{(4\pi)^2}
+ \frac{1}{4\pi}\, \sum_{l=1}^2 \sum_{m=-l}^{l}\, \alpha^{(1)}_{lm} \, Y_{lm}\big (\theta_{1}, \phi_{1}\big ) \nonumber \\[2mm]
& \phantom{xx}+ \frac{1}{4\pi}\, \sum_{l=1}^2 \sum_{m=-l}^{l}\, \alpha^{(2)}_{lm} \, Y_{lm}\big (\theta_{2},\phi_{2}\big )
\nonumber \\[2mm]
& \phantom{xx}+ \sum_{l=1}^2 \sum_{l^\prime=1}^2 \sum_{m=-l}^{l}\sum_{m^\prime=-l^\prime}^{l^\prime}\, \gamma_{lm l^\prime m^\prime} \, Y_{lm}\big (\theta_{1}, \phi_{1}\big ) \, Y_{l^\prime m^\prime}\big (\theta_{2}, \phi_{2}\big ) \,,
\end{align} 
with the decay angles computed w.r.t~some reference axis, the spin-quantisation axis, which in this work we choose to be
 the so-called \emph{modified helicity coordinate system} \cite{Aaboud:2019gxl}.\footnote{It defines the reference axis as the spatial direction of the boson $V_1$ in the centre-of-mass frame of the di-boson system $V_1V_2$. This choice implies that the polarisations and spin correlations are defined in the centre-of-mass frame of the di-boson system.}
Since  $Y_{lm}\big (\theta_{i}, \phi_{i}\big )$  are an orthogonal basis, it is possible to extract the $\alpha$ and $\gamma$ coefficients by projecting the angular distributions on them.

Since the analytic structure of both the decay density matrix $\Gamma$ and the spin-density matrix $\rho$ are known at LO, it is possible to extract from the $\alpha_{lm}$ and $\gamma_{lm l^\prime m^\prime}$ coefficients the entries of $\rho$.
In particular, the  QT approach relies on the LO-accurate $\PZ$-boson decay matrix \cite{Boudjema:2009fz,Rahaman:2021fcz},
\beq\label{eq:LOgamma}
\Gamma(\theta,\phi)\,=\,\frac14\left(\begin{array}{ccc}
     {1+2\eta_\ell\cos\theta+\cos^2\theta} & \sqrt{2}\,\sin\theta(\eta_\ell + \cos\theta)e^{\ri \phi}  & (1-\cos^2\theta) e^{2\ri \phi}\\
     \sqrt{2}\,\sin\theta(\eta_\ell + \cos\theta) e^{-\ri \phi}
     & 2\,({1-\cos^2\theta}) & \sqrt{2}\,\sin\theta(\eta_\ell - \cos\theta) e^{\ri \phi}\\
     (1-\cos^2\theta) e^{-2\ri \phi}& 
     \sqrt{2}\,\sin\theta(\eta_\ell - \cos\theta)e^{-\ri \phi}& {1-2\eta_\ell\cos\theta+\cos^2\theta}
\end{array}\right)\,.
\eeq
The quantity $\eta_\ell$ in Eq.~\eqref{eq:LOgamma} is the so called spin-analysing power,\footnote{The parameter $\eta_\ell$ in this work is equivalent to $-\alpha_\ell$ in \citere{DelGratta:2025qyp}.} which for massless charged leptons in $\PZ$ decays reads, at LO, 
\beq
\eta_\ell = \frac{2\,g_{V,\ell}\,g_{A,\ell}}{g_{V,\ell}^2+g_{A,\ell}^2}=\frac{1-4\sin^2\theta_{\rm w}}{1-4\sin^2\theta_{\rm w}+8\sin^4\theta_{\rm w}}\,= 0.2131\ldots\,,
\label{eq:etalLO}
\eeq
with the input parameter that are listed in Sec.~\ref{sec:decay} and with $g_{V,\ell},\,g_{A,\ell}$ being respectively the vector and axial-vector coupling of the $\PZ$ boson to leptons.
This parameter has been  widely investigated at LEP through asymmetries \cite{ALEPH:2005ab}, and as already mentioned in the introduction, it is the culprit of the giant EW corrections appearing in QT with $\PZ$ bosons that decay leptonically. It receives large
corrections since the $(1-4\sin^2\theta_{\rm w})$ factor is small due to an accidental cancellation.

In quantum mechanics the $\rho$ matrix for a two-qutrit system  can be conveniently written in terms of the irreducible tensorial representations of the boson spin \cite{Aguilar-Saavedra:2022wam},
\begin{equation}
\rho = \frac{1}{9} \left[\mathbf{1}_3  \otimes\mathbf{1}_3  + A^{(1)}_{lm} (T_{lm} \otimes \mathbf{1}_3 )+ A^{(2)}_{lm} (\mathbf{1}_3  \otimes T_{lm}) + C_{lml'm'} (T_{lm} \otimes T_{l'm'}) \right].
\label{eq:rhoexp}
\end{equation}
The different symmetries of the system determine which of the $A$ and $C$ coefficients are non-vanishing and the relations amongst them  \cite{DelGratta:2025qyp}. In particular, knowing all the  $A$ and $C$ coefficients, the whole $\rho$ matrix can be reconstructed.

Using the same basis, the decay matrix $\Gamma$ in Eq.~\eqref{eq:LOgamma} can also be rewritten as
\begin{equation}
\Gamma = \frac{1}{3} \left[\mathbf{1}_3   + \sqrt{2\pi} \, \eta_\ell \, T_{1m} Y_{1m} + \sqrt{\frac{2\pi}{5}} \, T_{2m} Y_{2m}  \right].
\label{eq:Gammaexp}
\end{equation}
Plugging Eqs.~\eqref{eq:rhoexp} and \eqref{eq:Gammaexp} into Eq.~\eqref{eq:RhoGammaGamma} it is possible to derive an expression of the same form of Eq.~\eqref{eq:VVang} (see \eg \citere{DelGratta:2025qyp} for more details) and derive relations between the coefficients $\alpha$, $\gamma$ and $A$, $C$:
\begin{align}\label{eq:gammas_to_Cs}
    &\sqrt{8\pi}\alpha_{1m} = \eta_\ell\, A_{1m},\qquad
    &\sqrt{40\pi}\alpha_{2m} = A_{2m},\qquad\qquad
    &8\pi\gamma_{1m1m'} = \eta_\ell^2 \,C_{1m1m'},\nnb\\
    &40\pi\gamma_{2m2m'} = C_{2m2m'},\qquad
    &8\pi\sqrt{5}\gamma_{2m1m'} = \eta_\ell \,C_{2m1m'},\qquad
    &8\pi\sqrt{5}\gamma_{1m2m'} = \eta_\ell \,C_{1m2m'}\,.
\end{align}

Thus, at LO and assuming NWA, it is possible via QT to extract $\alpha$ and $\gamma$ coefficients and, since $\Gamma$ is known, convert them to $A$ and $C$ coefficients and in turn the entries of $\rho$. The exact same results would be obtained via the HA, calculating directly $\rho_{\lambda_1,\lambda_1',\lambda_2,\lambda_2'}
\Gamma_{\lambda_1,\lambda_1'}(\theta_1,\phi_1)
\Gamma_{\lambda_2,\lambda_2'}(\theta_2,\phi_2)$ and removing/dividing by the contribution of the $\Gamma$'s.

\medskip 

Before moving on to the NLO case, we provide some more details on the HA approach. If for a $\PZ+X$ production process, with subsequent $\PZ$ decay, we denote as $\mc A_{\rm unp}$ the unpolarised amplitude associated to the production $\times$ decay mechanism,  this amplitude can be written as a sum of physical-helicity amplitudes \cite{Denner:2020bcz}:
\beq
\mc A_{\rm unp}\,=\,
\mc A^{\rm (prod)}_{\mu}\,
\frac{\sum_{\lambda} \varepsilon^{*\,\mu}_\lambda\varepsilon^{\phantom{*\,}\nu}_{\lambda}}{k_\PZ^2-M_\PZ^2+\ri M_\PZ \Gamma_\PZ}\,
\mc A^{\rm (dec)}_{\nu}
\,=\,
\sum_{\lambda}\,
\frac{ 
\mc A^{\rm (prod)}_{\lambda}
\mc A^{\rm (dec)}_{\lambda}
}{k_\PZ^2-M_\PZ^2+\ri M_\PZ \Gamma_\PZ}\,\equiv\,\sum_\lambda \mc A_{\lambda}\,,
\label{eq:amp_sum}
\eeq
where $k_\PZ$, $\MZ$ and $\GZ$ are the four momentum, the pole mass and the pole width of the $\PZ$ boson, respectively. 
In doing so we have just assumed the intermediate $\PZ$ boson in the $s$-channel and that $A^{\rm (prod)}$ and $A^{\rm (dec)}$ are evaluated with $k_{\PZ}^2=\MZ^2$,  as done in the pole approximation (PA) \cite{Ballestrero:2017bxn,Ballestrero:2019qoy,Ballestrero:2020qgv,Denner:2020bcz,Denner:2020eck,Poncelet:2021jmj,Denner:2021csi,Le:2022lrp,Le:2022ppa,Denner:2022riz,Dao:2023pkl,Pelliccioli:2023zpd,Denner:2023ehn,Dao:2023kwc,Denner:2024tlu,Dao:2024ffg,Haisch:2025jqr}.
The amplitude $\mc A_{\rm unp}$  can therefore be factorised, helicity by helicity, into a production amplitude $\mc A^{\rm (prod)}_{\lambda}$ and a decay amplitude $\mc  A^{\rm (dec)}_{\lambda}$, possibly taking into account off-shell effects in the denominator of the propagator. The NWA approximation is simply obtained in the $ \Gamma_\PZ/ M_\PZ\to 0$ limit, after $k_\PZ^2$ integration.\footnote{In the literature, the NWA  can also consists of a off-shell reshuffling of the amplitude evaluated on-shell \cite{Richardson:2001df,Artoisenet:2012st,BuarqueFranzosi:2019boy,Pellen:2021vpi,Hoppe:2023uux}. In this paper we will refer to NWA as strictly on-shell. We note that anyway, also taking into account off-sell effects, it is not equivalent to the pole expansion, see \eg \citere{Carrivale:2025mjy}.}

Squaring Eq.~\eqref{eq:amp_sum}, we obtain
\beq\label{eq:sqamp}
\left|\mc A_{\rm unp}\right|^2
\,=\,
\left|\sum_\lambda \mc A_{\lambda}\right|^2
\propto 
\sum_{\lambda,\lambda'} \left(\mc{A}^{\rm (prod)*}_\lambda \mc{A}^{\rm (prod)}_{\lambda'}\right) \left(\mc{A}^{\rm (dec)*}_\lambda \mc{A}^{\rm (dec)}_{\lambda'}  \right)
\eeq 
where in the NWA, besides normalisation constants, $\mc{A}^{\rm (prod)*}_\lambda \mc{A}^{\rm (prod)}_{\lambda'}$ and $\mc{A}^{\rm (dec)*}_\lambda \mc{A}^{\rm (dec)}_{\lambda'}$ respectively correspond to $\rho_{\lambda,\lambda'}$ and $\Gamma_{\lambda,\lambda'}$ in Eq.~\eqref{eq:RhoGamma}. 
More in detail, if we call $\sigma^{\rm prod}$ the unpolarised cross section for the production of the $\PZ+X$ process, so calculated via $\left|\sum_\lambda \mc{A}^{\rm (prod)}_{\lambda}\right|^2$, and instead  $\sigma^{\rm prod}_{\lambda,\lambda'}$ if calculated via $\left(\mc{A}^{\rm (prod)*}_\lambda \mc{A}^{\rm (prod)}_{\lambda'}\right)$, then
\beq
\rho_{\lambda,\lambda'}=\frac{\sigma^{\rm prod}_{\lambda,\lambda'}}{\sigma^{\rm prod}}\,. \label{eq:rhofromsigma}
\eeq
This definition can be easily extended in the case of multiple $\PZ$ bosons, as in the case of Eq.~\eqref{eq:RhoGammaGamma}.
Analogously, if we call $\sigma^{\rm dec}$ the unpolarised decay width for the $\PZ\to \ell^+\ell^-$, so calculated via $\left|\sum_\lambda \mc{A}^{\rm (dec)}_{\lambda}\right|^2$, and  $\sigma^{\rm dec}_{\lambda,\lambda'}$ when it is calculated via $\left(\mc{A}^{\rm (dec)*}_\lambda \mc{A}^{\rm (dec)}_{\lambda'}\right)$, then
\beq
\Gamma_{\lambda,\lambda'}=\frac{\sigma^{\rm dec}_{\lambda,\lambda'}}{\sigma^{\rm dec}}\,. \label{eq:Gammafromsigma}
\eeq

In conclusion, via the HA method, each term entering in the sum of the r.h.s.~of Eq.~\eqref{eq:sqamp} can be computed, and the $\rho_{\lambda,\lambda'}$ can be either calculated directly or extracted by the HA by simply dividing by the corresponding $\Gamma_{\lambda,\lambda'}$. This is particularly convenient when acceptance or selection cuts on the final states need to be taken into account, as done in Sections~\ref{sec:dress} and \ref{sec:ZZ}. 
The same applies to an arbitrary number of intermediate EW bosons. 

In general, the HA method can be used for the calculation of integrated and differential  polarised cross sections. In the case of $\PZ$-boson pairs, it has already been applied up to NLO EW + NNLO QCD accuracy as well as in the presence of parton-shower-matching and multi-jet-merging effects \cite{Carrivale:2025mjy}. 
In principle, each component $\mc{A}^{\rm (prod)}_{\lambda}$ and $\mc{A}^{\rm (dec)}_{\lambda}$ can be independently calculated. In practice, as already mentioned, most of the available MC tools evaluate $\mc{A}_{\lambda}$, such that cuts can directly be taken into account at the production $\times$ decay level.  

\subsection{Next-to-leading order}

As already discussed in the introduction, the extraction via QT of the coefficients of the spin-density matrix in the presence of leptonically decaying $\PZ$ bosons is plagued by large EW corrections \cite{Grossi:2024jae}. They  emerge from the  decay matrix $\Gamma$ and are erroneously propagated onto the spin-density matrix $\rho$ \cite{DelGratta:2025qyp} if in the relation between $\gamma_{1m1m'}$ and $C_{1m1m'}$ in Eq.~\eqref{eq:gammas_to_Cs} the LO value of the spin-analysing power $\eta_\ell$  is employed. In \citere{Goncalves:2025qem} it has been shown that by simply replacing in the definition of $\eta_\ell$ in Eq.~\eqref{eq:etalLO} the NLO EW corrected value of the EW mixing angle, $\sweff$, the difference between $C_{1m1m'}$ coefficients obtained at LO and NLO diminishes. Still, the consistency of this procedure has to be better understood and validated.

In the NWA or PA, via the HA method, Eq.~\eqref{eq:RhoGamma} can be calculated at NLO EW accuracy and, by construction, it preserves the factorised form that is present in the r.h.s.~of the equation. This means that both the decay matrix $\Gamma$ and the $\rho$ spin-density matrix can be separately calculated at NLO EW accuracy. Thus, it is possible to check under which conditions the extraction of the $\rho$ spin-density matrix via QT returns the same values obtained via the HA method. In other words, when the QT-extracted value corresponds to the ``true'' value.  In practice, in the case of QT, the NLO EW corrections are evaluated to the full process of production and decay. The requirement to use the PA or the NWA is not present and also non-resonant effects can be calculated. With QT, the coefficients $\alpha$ and $\gamma$  are directly calculated via the momenta of the final-state leptons  and converted to $A$ and $C$  and in turn to the entries of $\rho$. We stress that  a specific assumption on the value of $\eta_\ell$ has to be taken for performing such  conversion. On the contrary, via the HA method the workflow is the opposite. The entries of $\rho$ are directly calculated, the $A$ and $C$ coefficients are derived from them and then converted to the $\alpha$ and $\gamma$ ones via the computed value of $\eta_\ell$.

The HA method directly gives access to the spin-density matrix of the $\PZ+X$ system, but it is a purely calculation method and by construction cannot account for contributions that do not feature  intermediate resonating $\PZ$ boson, which instead can be present for the final state $\ell^+\ell^- + X$. On the contrary, the QT is applied directly on the $\ell^+\ell^- + X$ final state, and therefore the NLO EW calculation can be performed in any approximation, however in the extraction of the $A$ and $C$ coefficients the intermediate $\PZ$ must be assumed and the value of $\eta_\ell$ must also be chosen properly. The comparison of the two methods precisely helps in this choice, and at the same time allows one to investigate effects related to non-resonant contributions or other features of the set-up considered, such as cuts. Especially in the context of the SM Higgs boson decay into four leptons, we will see in Sec.~\ref{sec:HZZ} the relevance of such comparison.

It is worth to remind the reader that so far the QT has not been exploited at the experimental level to extract the complete spin-density matrix in diboson final states from LHC data. However, this technique has been applied by the CMS Collaboration in top-quark pair final states to determine the corresponding spin density matrix elements~\cite{CMS:2024zkc}. 
In the case of diboson processes, the polarisation fractions have been extracted from LHC Run-2 data via polarisation-template fits \cite{Aaboud:2019gxl,CMS:2020etf,CMS:2021icx,ATLAS:2022oge,ATLAS:2023zrv,ATLAS:2024qbd,ATLAS:2025wuw}. These experimental results rely on helicity-dependent theoretical predictions (polarisation templates) obtained with the HA method tailored to the 9 diagonal (azimuthal-averaged) entries of the spin-density matrix \cite{Ballestrero:2017bxn,BuarqueFranzosi:2019boy,Denner:2020bcz,Poncelet:2021jmj,Denner:2021csi,Le:2022lrp,Hoppe:2023uux,Pelliccioli:2023zpd,Denner:2023ehn,Dao:2023kwc,Denner:2024tlu,Carrivale:2025mjy,Haisch:2025jqr}.

\medskip 

To ensure a consistent comparison between QT and HA results, it should be emphasised that the calculation of NLO EW corrections to a cross section includes only the exact $\mathcal{O}(\alpha_{\rm ew})$ contributions. Higher-order terms, such as $\mathcal{O}(\alpha_{\rm ew}^2)$ and beyond, are not taken into account. On the other hand, the $\rho$ and the $\Gamma$ must respectively satisfy the conditions ${\rm Tr} (\rho) =1$ and ${\rm Tr} (\Gamma )=1$, such that both at LO and NLO EW accuracy one has to keep $\mathcal{O}(\alpha_{\rm ew})$ corrections both in the numerator and the denominator in their definitions (Eqs.~\eqref{eq:rhofromsigma} and \eqref{eq:Gammafromsigma}), with no exact expansion. This means that, \eg, if we consider Eq.~\eqref{eq:RhoGammaGamma}, which we rewrite in a compact way as
\beq
\label{eq:RhoGammaGammaLO}
\frac{16\pi^2}9\frac1{\sigma^{\rm LO}}\frac{\rd^4\sigma^{\rm LO}}{\rd\Omega_1\,\rd\Omega_2}\,=\,
\rho^{\rm LO}\,
\Gamma_1^{\rm LO}\,
\Gamma_2^{\rm LO}\,,
\eeq
at NLO EW accuracy, assuming the same factorised structure that is present in Eq.~\eqref{eq:amp_sum},  one obtains
\begin{eqnarray}
\nonumber
\frac{16\pi^2}9\frac1{\sigma^{\rm NLO}}\frac{\rd^4\sigma^{\rm NLO}}{\rd\Omega_1\,\rd\Omega_2}\,&=&\,
\rho^{\rm LO}\,
\Gamma_1^{\rm LO}\,
\Gamma_2^{\rm LO}\,\nnb\\[0.05cm]
&+&
\delta\rho^{\rm NLO}\,
\Gamma_1^{\rm LO}\,
\Gamma_2^{\rm LO}\,+
\rho^{\rm LO}\,
\delta\Gamma_1^{\rm NLO}\,
\Gamma_2^{\rm LO}\,+
\rho^{\rm LO}\,
\Gamma_1^{\rm LO}\,
\delta\Gamma_2^{\rm NLO}\,\\[0.2cm]
\label{eq:RhoGammaGammaNLO}
&+&\mathcal{O}(\alpha^2_{\rm ew})\,,\nnb
\end{eqnarray}
where  $\delta \rho^{\rm NLO}$ and $\delta \Gamma_i^{\rm NLO}$ are defined through
\beq
\rho^{\rm NLO}= \rho^{\rm LO}+\delta \rho^{\rm NLO}+\mathcal{O}(\alpha^2_{\rm ew})\,,\qquad
  \Gamma_i^{\rm NLO}=\Gamma_i^{\rm LO}+\delta \Gamma_i^{\rm NLO}+\mathcal{O}(\alpha^2_{\rm ew})\,.
  \eeq
 In all the previous equations we have dropped the spin indices and we imply the summation over them. 
  The equations are valid both for NWA and double-pole-approximation (DPA), since the additional Breit-Wigner that is present in the latter it is exactly canceled by the $1/\sigma^{\rm NLO}$ normalisation.
   We want to stress that the $\mathcal{O}(\alpha^2_{\rm ew})$ term appearing in Eq.~\eqref{eq:RhoGammaGammaNLO} is not related to simultaneous $\mathcal{O}(\alpha_{\rm ew})$ corrections in the production and/or one or more decay entering $\sigma$. Such contributions are never present in an NLO EW calculation for the cross section of the production + decay process; they would appear only at NNLO EW or at higher orders.\footnote{The relevance of these contributions in the comparison with data is precisely the subject of Sec.~\ref{sec:pseudodata}. } Instead, the $\mathcal{O}(\alpha^2_{\rm ew})$ term corresponds to contributions that involve at the same time, after the expansion, corrections in the production and in one decay (or in two different decays), but with one contribution coming from the numerator of the  l.h.s.~of Eq.~\eqref{eq:RhoGammaGammaNLO} and the other one from the denominator. In other words, they depend on the unpolarised production and/or decay. However,  if at the unpolarised and inclusive level the NLO corrections are small, as typically it is unless very boosted kinematics are selected, such contributions are negligible in our discussion.  

 Following the same logic, when $\eta_{\ell}$ is present, the relations that allow for translating the information from angular distributions ($\alpha$ and $\gamma$) to the coefficients of the $\rho$ matrix ($A$ and $C$), see Eq.~\eqref{eq:gammas_to_Cs}, are expected to be modified as
 \begin{eqnarray}\label{eq:gammas_to_Cs_NLO}
\sqrt{8\pi}\alpha_{1m}^{\rm NLO} &=& \eta_\ell^{\rm LO} \,A_{1m}^{\rm LO}\left(1+ \frac{\delta \eta^{\rm NLO}_\ell}{\eta^{\rm LO}_\ell}+\,\frac{\delta A_{1m}^{\rm NLO}}{A_{1m}^{\rm LO}}\right)\,, \qquad\\
8\pi\sqrt{5}\gamma_{1m2m'}^{\rm NLO} &=& \eta_\ell^{\rm LO} \,C_{1m2m'}^{\rm LO}\left(1+ \frac{\delta \eta^{\rm NLO}_\ell}{\eta^{\rm LO}_\ell}+\,\frac{\delta C_{1m2m'}^{\rm NLO}}{C_{1m2m'}^{\rm LO}}\right)\,, \qquad\\
    8\pi\gamma^{\rm NLO}_{1m1m'} &=& (\eta^{\rm LO}_\ell)^2 \,C_{1m1m'}^{\rm LO}\left(1+2 \frac{\delta \eta^{\rm NLO}_\ell}{\eta^{\rm LO}_\ell}+\,\frac{\delta C_{1m1m'}^{\rm NLO}}{C_{1m1m'}^{\rm LO}}\right)\,, \label{eq:gammatoCNLO}
\end{eqnarray}
where the same convention already adopted for $\rho $ and $\Gamma$ has been used in order to denote LO and NLO quantities also for $\eta_\ell$ and the $A$ and $C$ coefficients.
The main point is that the NLO EW version of each quantity  cannot be simply plugged into Eq.~\eqref{eq:RhoGammaGammaNLO}  for a consistent comparison between the results obtained via QT or the HA method.

A crucial step is also the definition of $\eta_\ell^{\rm NLO}$. In order to correctly retain the relation between  the $\alpha,\gamma$ and $A,C$ coefficients,  $\eta_\ell^{\rm NLO}$ has to be properly defined. One needs that also at NLO EW accuracy the analytic formula of Eq.~\eqref{eq:LOgamma} is preserved and that, if it is the case,  $\eta_\ell^{\rm NLO}$ is the corresponding value of $\eta_\ell$. This fact is far from obvious, since the NLO EW corrections also include the real radiation of photons ($\PZ\to \ell^+ \ell^-\gamma$). The decay matrix could have higher powers in $\cos \theta$ or $e^{i\phi}$, {\it i.e.}, it could depend on  $Y_{lm}\big (\theta, \phi \big )$ with $l>2$. 

We stress that this definition of $\eta_\ell$, which we dub $\eta_\ell^{\rm NLO}$, is different form taking the LO definition in Eq.~\eqref{eq:etalLO} and directly plug the value $\sweff$:
\beq
\label{eq:etaleff}
\eta_\ell^{\rm eff}\equiv \eta^{\rm LO}_\ell\Big|_{\sin^2\theta_{\rm w} \to \sweff}= 
\frac{1-4\sweff}{1-4\sweff+8\sin^4\!\theta^{\rm \,eff}_{\rm w}}\,.
\eeq 
Still, we reckon that $\sweff$ has been precisely defined for the asymmetries on the $\PZ$ resonance and so it is expected that if the analytic structure of the $\Gamma$ matrix is unaltered at NLO, the two definitions,  
$\eta_\ell^{\rm eff}$ and $\eta_\ell^{\rm NLO}$, should be equivalent. This is precisely what we investigate in Sec.~\ref{sec:etaNLO}.
First in Sec.~\ref{sec:decay} we calculate the decay density matrix $\Gamma$ at NLO EW accuracy and we check the validity of the approximating it via a LO simulation with $\eta_\ell^{\rm eff}$.
Then in Sec.~\ref{sec:dress} we evaluate it, via HA, in realistic processes (production $\times$ decay), taking into account cuts in the laboratory frame.

\section{Spin-analysing power at NLO EW}
\label{sec:etaNLO}

\subsection{$\PZ\to \ell^+ \ell^-$ and the decay matrix}\label{sec:decay}

\begin{figure*}
  \centering
  \includegraphics[width=0.24\textwidth]{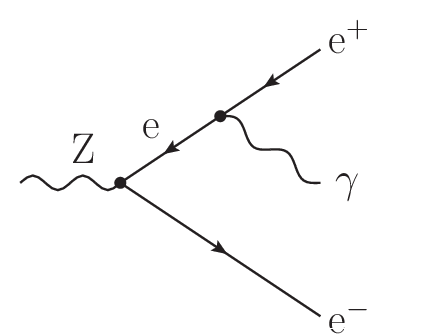}
  \includegraphics[width=0.24\textwidth]{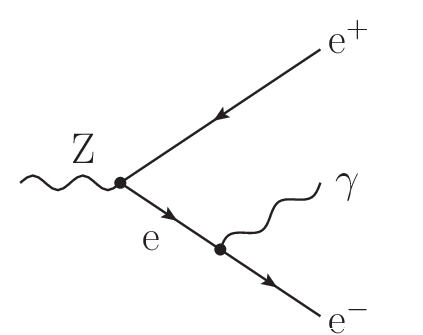}\\
  \includegraphics[width=0.24\textwidth]{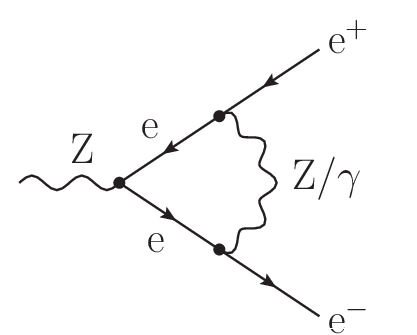}
  \includegraphics[width=0.24\textwidth]{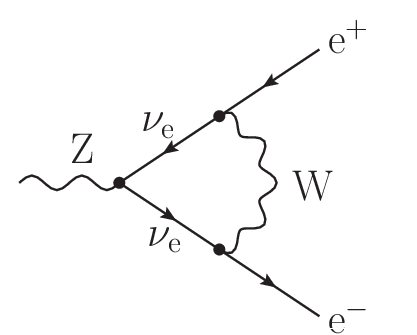}
  \includegraphics[width=0.24\textwidth]{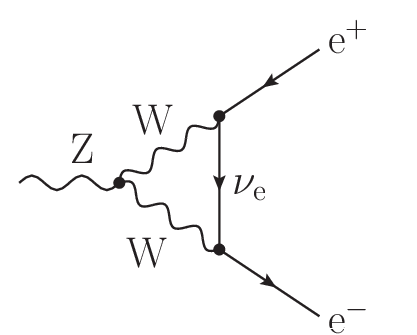}
  \caption{Real (top row) and virtual (bottom row) diagrams contributing to the Z-boson decay into a massless lepton--antilepton pair at NLO EW.
  }\label{fig:NLOdiags}
\end{figure*}

We have applied the HA method introduced in Sec.~\ref{sec:QTvsHA} to carry out the calculation of the 8 independent entries  of the NLO EW decay matrix, $\Gamma^{\rm NLO}$, for the inclusive process
$ \PZ\,\to\ell^+\ell^-\,(+X)$, which includes also the contributions from the real emission of photons  $ \PZ\,\to\ell^+\ell^-\,\gamma$. Such contributions could in principle disrupt the two-body nature of the analytic structure of the
entries of Eq.~\eqref{eq:LOgamma}.
Representative diagrams entering the calculation, both for real and virtual contributions are shown in Fig.~\ref{fig:NLOdiags}.
The calculation has been performed with two fully independent codes: 
\begin{itemize}
    \item (MG5) a private version of {\sc MadGraph5\_aMC@NLO} \cite{Alwall:2014hca,Frederix:2018nkq}, version 3.3.2, which allows for computing diagonal entries of the decay matrix at NLO EW accuracy.
    \item (RCL) a standalone MC code based on \recola1 \cite{Actis:2016mpe} amplitudes (version 1.4.4), \collier \cite{Denner:2016kdg} loop integrals (version 1.2.7), a modified implementation of the {\sc Vegas} numerical-integration algorithm \cite{Baglio:2024gyp}, and the abelianisation of the FKS subtraction scheme for QED IR singularities \cite{Frixione:1995ms}, allowing to compute both diagonal and off-diagonal entries of the decay matrix at NLO EW accuracy.
\end{itemize}
Excellent agreement was found at integrated and differential level, for the diagonal entries of the decay matrix,
providing a strong validation of the results presented in this section.

\medskip

The numerical results correspond to the following choices for the EW-boson on-shell masses
\cite{ParticleDataGroup:2024cfk},
\begin{alignat}{2}\label{eq:ewmasses}
  \MWOS &= 80.377 \GeV,\qquad \MZOS &= 91.1876 \GeV \,.
\end{alignat}
The calculation is performed in the $G_\mu$ scheme \cite{Sirlin:1980nh,Denner:2000bj},
 with the Fermi constant set to $\GF = 1.16638\cdot 10^{-5} \GeV^{-2}$.
The masses of the top quark and of the Higgs boson are set equal to \cite{ParticleDataGroup:2024cfk},
\begin{alignat}{2}\label{eq:THmasses}
  \Mt &= 172.69 \GeV,&\qquad \MH &= 125.25\GeV\,.
\end{alignat}
The NLO partial decay width for the considered decay (for a single lepton family) features very good agreement between 
the two independent codes:
\begin{align}
\Gamma_{\ell^+\ell^-}^{\rm (MG5)} = 0.08383(3) \GeV,\qquad
\Gamma_{\ell^+\ell^-}^{\rm (RCL)} = 0.083834(4) \GeV\,. \nnb    
\end{align}

We have computed all entries of the decay matrix at LO, at exact NLO EW accuracy, and in the EW-virtual approximation \cite{Kallweit:2015dum}.\footnote{
In this approximation one-loop corrections are included after being IR regulated through the EW generalisation of the Catani operator \cite{Dittmaier:1999mb}, while hard-real photon contributions are discarded. 
\label{foot:approx}
} 
In all cases the computation is performed differentially in the polar and azimuthal angle of the positively charged lepton. Since we treat the leptons as massless,  in the case of exact NLO EW corrections leptons are clustered together with radiated photons  (dressed leptons) if 
\beq
\Delta R_{\ell\gamma}\,=\,\sqrt{(y_{\ell}-y_{\gamma})^2+(\phi_{\ell}-\phi_{\gamma})^2} < R\,,
\eeq
with $R$ corresponding to the resolution radius.  

\begin{figure*}
  \centering
  \includegraphics[width=0.99\textwidth]{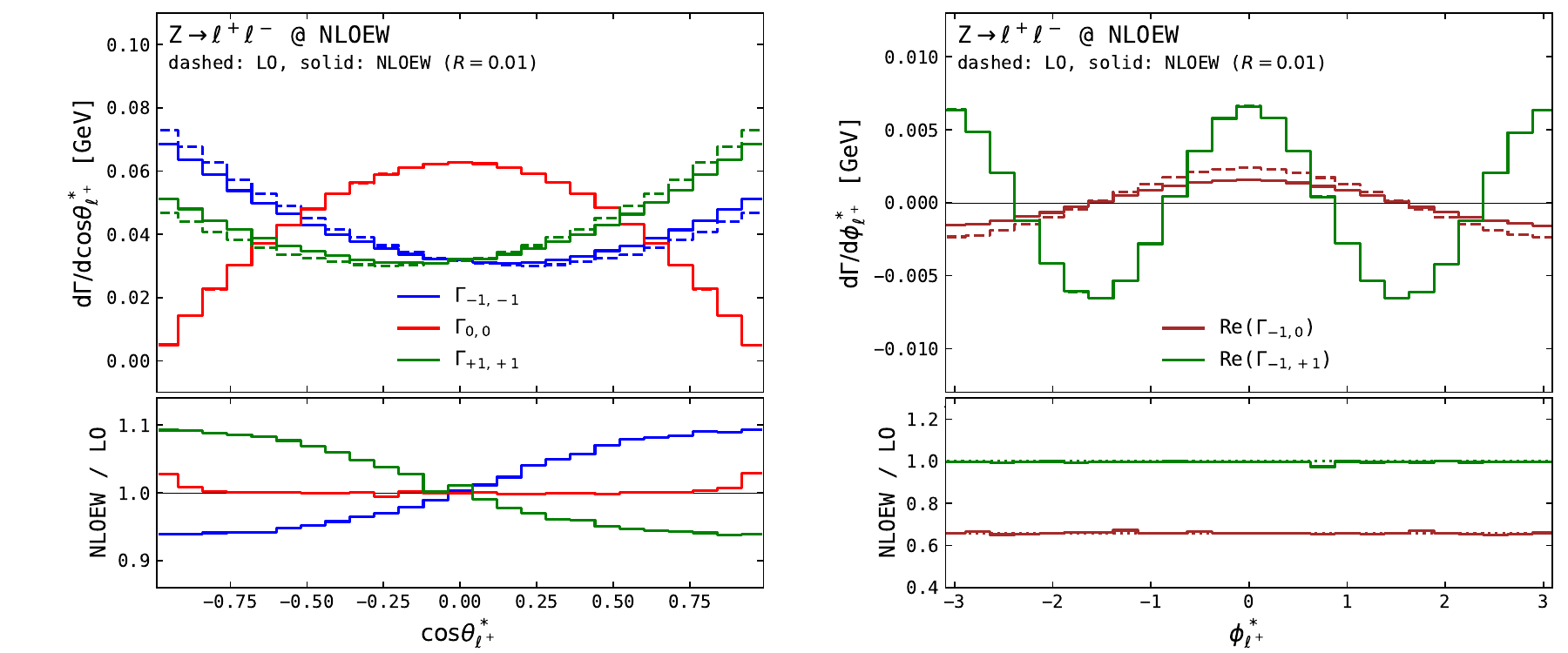}
  \caption{Z-boson decay into a massless lepton--antilepton pair, obtained at LO and NLO EW in the $G_\mu$ EW scheme and for $R=0.01$ dressing resolution. 
  Left figure: diagonal entries of the decay matrix $\Gamma$, differentially in the polar decay angle. Right figure: real part of two off-diagonal entries, differentially in the azimuthal decay angle.
  All results have been obtained through a dummy MC code based on \recola1, version 1.4.4 \cite{Actis:2016mpe}, a modified implementation of the {\sc Vegas} numerical-integration algorithm \cite{Baglio:2024gyp}, and the FKS subtraction scheme for QED IR singularities \cite{Frixione:1995ms}.
  }\label{fig:NLOfinal}
\end{figure*}

In Fig.~\ref{fig:NLOfinal} we show differential distributions at LO (dashed) and NLO EW (solid) accuracy in the polar and azimuthal decay angle of the dressed lepton $\ell^+$,  obtained with our direct computation of the $\Gamma^{\rm NLO}$ entries via HA.
The angle is in the $\PZ$-boson rest frame and we find that results obtained for $R=\,0.01,\,0.1,\,1$ agree with each other within integration uncertainties. We therefore show only results for the choice  $R=\,0.1$.

In the left panel the polar decay angle is considered, for the three diagonal entries: left-left $(-1,-1)$, longitudinal-longitudinal $(0,0)$, and right-right $(+1,+1)$. At LO, the three diagonal entries have a quadratic  and linear dependence on the cosine of the polar angle, as can be evinced from Eq.~\eqref{eq:LOgamma}. For the two entries $\Gamma_{\pm 1,\pm 1}$, corresponding to the transverse helicity of the $\PZ$ boson, the  quadratic dependence on $\cos\theta_{\ell^+}$ is the same at LO and at NLO, as can be understood by the ratio between NLO EW and LO predictions shown in the inset. Instead, the linear term, proportional to $\eta_\ell$, changes sizeably between LO and NLO EW. Performing a quadratic fit of the curves we obtain the values 
\begin{align}
\eta^{\rm LO}_\ell = 0.2131(1)\,,\qquad
\eta^{\rm EW,virt}_\ell = 0.1409(1)\,,\qquad
\eta^{\rm NLO}_\ell = 0.1405(8)\,,
\label{eq:etalNLO}    
\end{align}
where the number in parentheses is the error on the last digit. 

We have checked numerically that the analytic form of the decay matrix at NLO EW can be assumed to be equal to the one at the LO.\footnote{Besides performing a fit, we have projected directly on the spherical harmonics the results obtained and verified that, within numerical errors, the structure is the one shown in Eq.~\eqref{eq:Gammaexp}.  } Only the value $\eta_\ell$ is modified,  receiving  corrections of order $-35\%$.
Such corrections are almost entirely of virtual origin, since the result does not depend on $R$ and is consistent with the one obtained in the approximation\footnote{See footnote \ref{foot:approx} for more details.} where the real photon emissions are neglected. This outcome, which was not evident a priori and required explicit verification, shows that our numerical evaluation of $\eta_\ell^{\rm NLO}$ agrees very well with the value of $\eta_\ell^{\rm eff}$ (see Eq.~\eqref{eq:etaleff}) that has been calculated in \citere{Goncalves:2025qem} with the publicly available {\sc Griffin} package \cite{Chen:2022dow}.
Strikingly, the three-body contributions from real-photon radiation have almost no effects in modifying the leading two-body structure of the decay for a $\PZ$ boson at rest. We will see later in the section that the presence of cuts can change considerably this statement.

The only deviation, which is tiny, from the analytic form that is present at LO  is found at the endpoints of the polar-angle distribution for the longitudinal entry $\Gamma_{0,0}$, where a $3\%$ enhancement is found, mostly coming from real-photon corrections and with a mild dependence on the dressing radius (values between $R=0.01$ and $R=1.5$ give rather similar results to the angular dependence of $\Gamma_{0,0}$).
However, owing to the LO suppression at $\cos\theta=\pm 1$, a quadratic fit of the $\Gamma^{\rm NLO}_{0,0}$ distribution gives a result compatible with the LO one within numerical and fit uncertainties. Still, these tiny deviations are a signal of contributions that depend on a higher power of $\cos\theta$, but that numerically have no impact and can be neglected.

In the right panel of Fig.~\ref{fig:NLOfinal} the azimuthal angle distribution is considered for the real part of the left-longitudinal and left-right entries of the decay matrix, {\it i.e.}, $\Gamma_{-1,0}$ and $\Gamma_{-1,+1}$. As expected from Eq.~\eqref{eq:LOgamma}, a modulation in $\cos\,2\phi_{\ell^+}$ and $\cos\,\phi_{\ell^+}$ is present for the two terms respectively, and this holds for the NLO results as well. The only difference between LO and NLO is the different amplitude of the $\cos\,\phi_{\ell^+}$ modulation in the $\Gamma_{-1,0}$ term. In this case, the ratio between NLO and LO results is flat and corresponds to the $\eta_\ell^{\rm NLO}/\eta_\ell^{\rm LO}\simeq 0.66$. Indeed,  $\Gamma_{-1,+1}$ does not depend on $\eta_\ell$, while, after integration over $\theta$, $\Gamma_{-1,0}\propto \eta_\ell$. 
On top of probing the analytic structure of the off-diagonal entries of the decay matrix, this represents a consistency check of our extraction of $\eta_\ell$ at NLO.
Although it is not shown in Fig.~\ref{fig:NLOfinal}, we have checked that the NLO EW corrections do not change the analytic dependence on $\cos\theta_{\ell^+}$ and $\phi_{\ell^+}$ of the imaginary parts of the off-diagonal entries of the decay matrix.

With this calculation, we have excluded further modifications at NLO accuracy to the $\PZ$-boson decay matrix, apart from the strikingly large corrections to the value of $\eta_\ell$. However, in realistic computations, including also a production mechanism for the $\PZ$ boson, additional complications arise. 

First, while in the calculation for an on-shell $\PZ$-boson decay relies on the on-shell masses, in a full-fledged calculation including the production and the decay of EW bosons, the input masses must be the pole ones, converted from the values in Eq.~\eqref{eq:ewmasses} through the relations of \citere{Bardin:1988xt}. This holds for calculations performed either in the PA \cite{Denner:2000bj,Denner:2005fg,Denner:2019vbn} or in the NWA \cite{Uhlemann:2008pm,Artoisenet:2012st}, in order to correctly capture partial off-shell effects associated to the intermediate $s$-channel bosons. 
The NLO values in Eq.~\eqref{eq:etalNLO} computed with EW-boson pole masses rather than with on-shell masses are very similar. We  found that differences in the two cases are at the per mille level. Thus, this effect can be safely neglected.

\begin{figure*}[!t]
  \centering
  \includegraphics[width=0.64\textwidth]{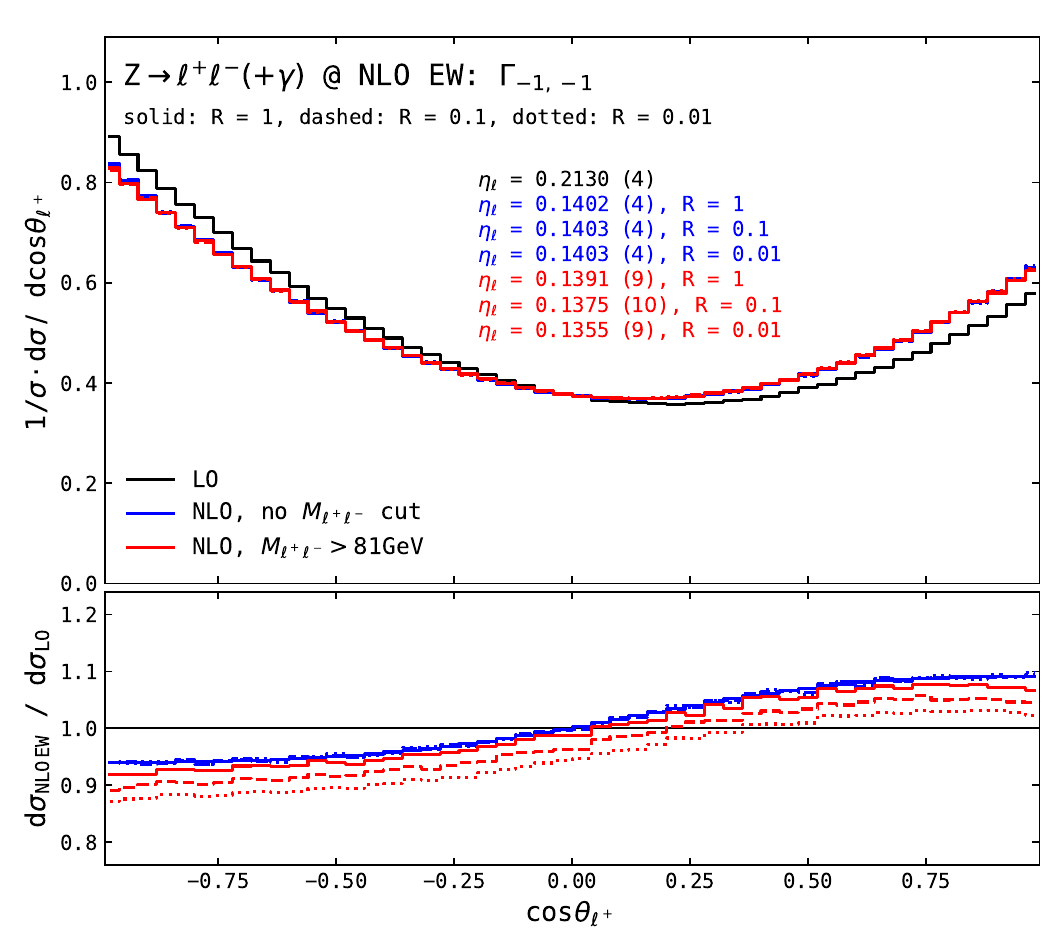}
  \caption{
  Z-boson decay into a massless lepton--antilepton pair, obtained at LO and NLO EW in the $G_\mu$ EW scheme and for various dressing-resolution radii. 
  The entry $\Gamma_{-1,-1}$ of the decay matrix is considered differentially in the polar decay angle. 
  All results have been obtained with a private version of MG5\_aMC@NLO \cite{Alwall:2014hca,Frederix:2018nkq} that enables the calculation of diagonal entries of the decay matrix at NLO accuracy. 
  }
  \label{fig:NLOdeccut}
\end{figure*} 

Second, at the experimental level, the reference frame of the $\PZ$ is defined as the one given by the momenta of the $\ell^+\ell^-$ pair. In an NLO EW calculation, for the Born, virtual and real emission contributions with photons clustered to leptons, the $\ell^+\ell^-$ pair and $\PZ$ rest frames are the same.  Instead, for the real emission contributions with non-clustered photons, the two frames are different. We have verified that the angular distributions and the value of $\eta^{\rm NLO}_\ell$ extracted in the two frames are equivalent and do not depend on $R$. However, in the experimental analyses $\ell^+\ell^-$ pairs are also required to  have an invariant mass close to  $\MZOS$, in order to be identified as a $\PZ$ boson. 

In Fig.~\ref{fig:NLOdeccut} we focus on the case of $\Gamma_{- 1,- 1}$ and we show the polar-angle distribution ($\cos \theta$) at LO together with the NLO EW prediction with the cut $M_{\ell^+\ell^-}> 81$ GeV (red)  and no cuts (blue), for three different $R$ choices: $R=1$ (solid), $R=0.1$ (dashed) and $R=0.01$ (dotted). Unlike the case with no cuts, there is a dependence on $R$ and, especially, the extracted value for $\eta_\ell^{\rm NLO}$ is different. The results are listed for the three cases in Tab.~\ref{Tab:etacuts}. 
\begin{table}[h!]
\begin{center}
\renewcommand{\arraystretch}{1.5}
\begin{tabular}{c|cccccc}
$\eta_\ell^{\rm NLO}$ & $R=1$ & $R=0.1$ & $R=0.01$ \\
\hline
$M_{\ell^+\ell^-}> 81$ GeV & 0.1391(9)  & 0.1375(10) & 0.1355(9) \\ [-0.2cm]
no $M_{\ell^+\ell^-}$ cut & 0.1402(4)  & 0.1403(4) & 0.1403(4) \\ 
\end{tabular}
\caption {
$\PZ$-boson spin-analysing power extracted from the polar-angle distribution of the $\Gamma_{-1,-1}$ entry of the decay matrix for three values of the lepton-dressing resolution radius. The invariant mass of the dressed-lepton pair is either uncut or constrained to be larger than $81\GeV$.
\label{Tab:etacuts}}
\end{center}
\end{table}
Although the values without the $M_{\ell^+\ell^-}$ cut are perfectly compatible with those of Eq.~\eqref{eq:etalNLO}, the differences between the values obtained for $M_{\ell^+\ell^-}>81\GeV$ and those of Eq.~\eqref{eq:etalNLO} are manifest. They consist of effects of order (at most) 5\%, which are definitely much smaller than relative difference between $\eta_\ell^{\rm NLO}$ and $\eta_\ell^{\rm LO}$, but they are not negligible. Furthermore, at the LHC a $\PZ$ boson is very rarely  produced at rest (in the laboratory frame). Since in a realistic set-up the recombination is defined in the laboratory frame, the effective value $R$ in the $\PZ$ boson rest frame depends on the momenta of the $\PZ$ boson itself. This aspect is investigated in the next section. 

\subsection{$\PZ$-boson decay in association to the production} \label{sec:dress}
The analysis of Sec.~\ref{sec:decay} assumes a $\PZ$ boson at rest, while in a LHC production processes a $\PZ$ boson is produced with a non-vanishing transverse momentum, at a given rapidity, and in association with other particles.
In order to study the impact of this different kinematic configuration  on the extraction of the spin-analysing power, we consider two specific processes, namely,  the inclusive production of a $\PZ$ boson in association with either another $\PZ$ boson or a $\PW$ boson at the LHC, with 13 TeV of centre-of-mass energy:
\beq\label{eq:WZZZ}
\Pp\Pp\rightarrow \PZ (\rightarrow \Pe^+\Pe^-)\,\PZ(\rightarrow\mu^+\mu^-),\qquad
\Pp\Pp\rightarrow \PW^+ (\rightarrow \Pe^+\nu_{\Pe})\,\PZ(\rightarrow\mu^+\mu^-)\,.
\eeq
We perform the calculation at NLO EW in the DPA with the {\sc MoCaNLO} general-purpose MC code \cite{Denner:2021csi,Grossi:2024jae,Carrivale:2025mjy}, which is interfaced with \recola1, version 1.4.4 \cite{Actis:2016mpe}, and relies on the dipole subtraction scheme for QED IR singularities \cite{Catani:1996vz,Catani:2002hc,Dittmaier:1999mb}.

We stress that the interplay between the dressing algorithm and the 
decay angles is not trivial, as at the LHC the dressing of charged leptons is applied to momenta evaluated in the laboratory frame, while the angular observables are constructed in the rest frame of target EW bosons, with respect to a certain axis of quantisation of the boson spins. For inclusive $\PZ\PZ$ and $\PZ\PW$ production it is rather natural \cite{Aaboud:2019gxl,Grossi:2024jae} to define this axis as the single-boson spatial direction in the boson-pair centre-of-mass frame.

\begin{figure*}
  \centering
  \includegraphics[width=0.46\textwidth]{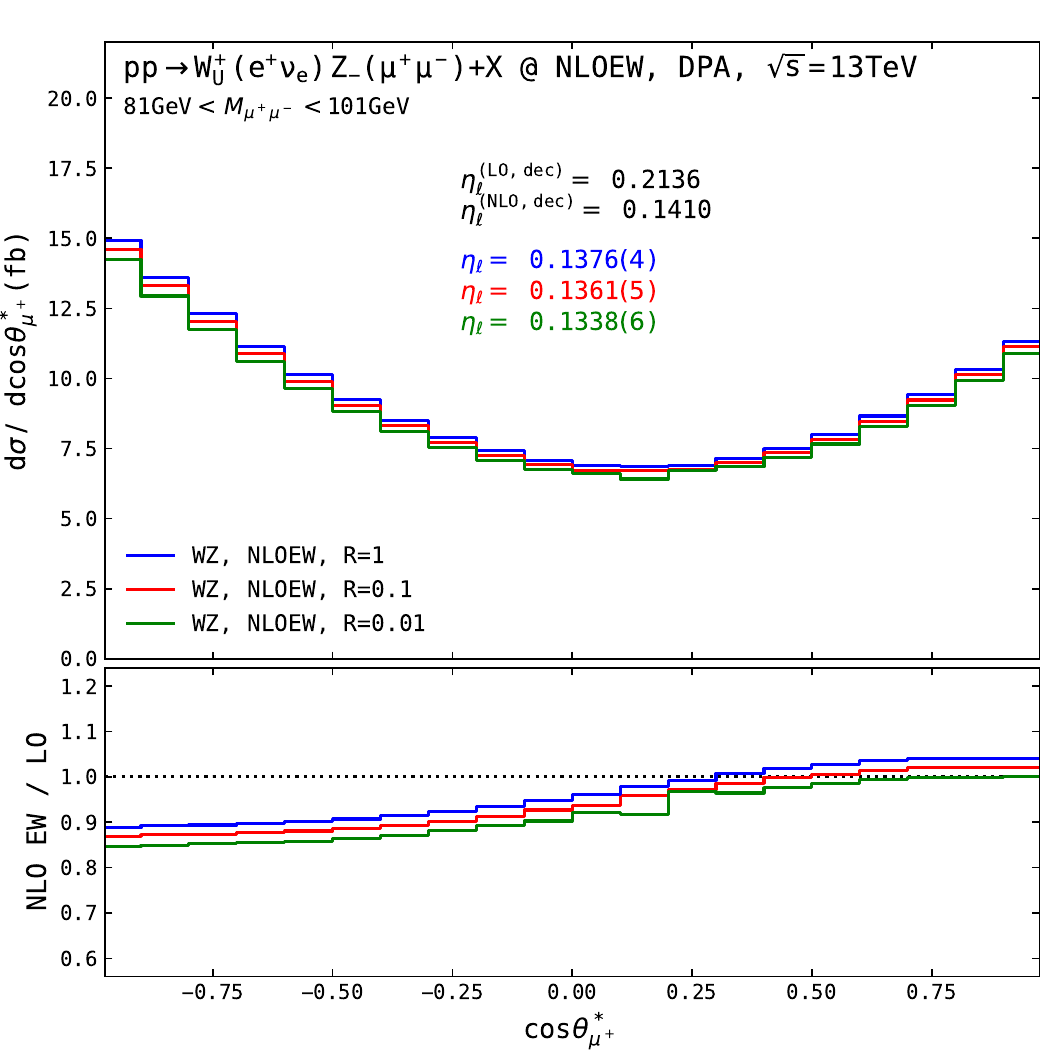}\qquad
  \includegraphics[width=0.46\textwidth]{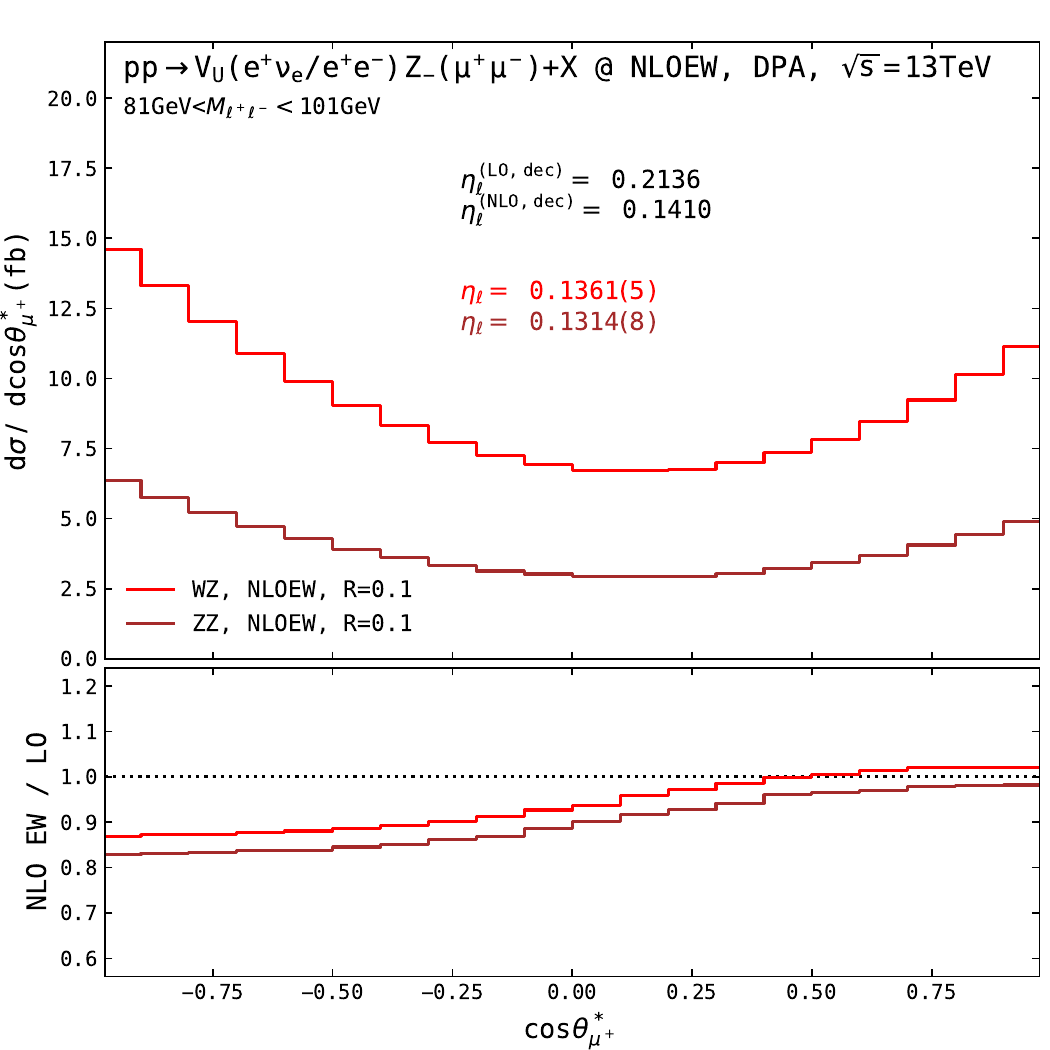}
  \caption{
Dependence of polar-angle distributions on the lepton-dressing radius (left) and on the production mechanism (right) for a left-handed $\PZ$ boson.
Left figure: left-handed $\PZ$ boson produced in association with an unpolarised $\PW$ boson, using different resolution radii in the lepton-dressing algorithm. Right figure: left-handed $\PZ$ produced with an unpolarised $\PZ$ boson (brown) or with an unpolarised $\PW$ boson (red).
An invariant-mass constraint $81\GeV<M_{\ell^+\ell^-}<101\GeV$ is applied to opposite-sign same-flavour lepton pairs, and an LHC collision energy of 13 TeV is understood.
All results have been obtained with the {\sc MoCaNLO} code \cite{Denner:2021csi,Grossi:2024jae,Carrivale:2025mjy}. }
\label{fig:etalNLO_WZ_ZZ}
\end{figure*}
In the left plot of Fig.~\ref{fig:etalNLO_WZ_ZZ} we show the distributions in the polar decay angle associated to a left-handed $\PZ$ boson that is produced together with an unpolarised $\PW^+$ boson and that is decaying into leptons.
This is equivalent to selecting in Eq.~\eqref{eq:amp_sum} only the $\lambda=-1$, such that in this case Eq.~\eqref{eq:RhoGamma} can be simplified as 
\beq\label{eq:RhoGammaLL}
\frac{4\pi}3\frac1{\sigma^{\rm (N)LO}}\frac{\rd^2\sigma^{\rm (N)LO}}{\rd\cos\theta\,\rd\phi}\,=\,
\Gamma^{\rm (N)LO}_{-1,-1}\,,
\eeq
where we have stressed that the relation is valid both at LO and NLO accuracy. Since the $\lambda=-1$ polarisation is chosen, the spin-density matrix is calculated directly on the only spin-state available so the $\rho$ has only one entry and therefore by definition it is equal to one.

We  repeat in the following the analysis of Sec.~\ref{sec:decay}. In this case we require 
\begin{equation}
81~{\rm GeV} < M_{\mu^+\mu^-} < 101~{\rm GeV}\,, \label{eq:Mmmcut}
\end{equation}
and apply the  standard cone dressing algorithm as  applied in ATLAS and CMS spin-correlation and polarisation analyses \cite{Aaboud:2019gxl,CMS:2020etf,ATLAS:2022oge,ATLAS:2023zrv,ATLAS:2024qbd,ATLAS:2025wuw}, consistently to what has been done in Sec.~\ref{sec:decay}.

Both at the inclusive and differential level  NLO EW corrections depend on the \sloppy{ recombination-radius} $R$, as can be seen in the left plot of Fig.~\ref{fig:etalNLO_WZ_ZZ}. Thus, also the extracted value for $\eta_\ell^{\rm NLO}$ depends on it, and it is listed in Tab.~\ref{Tab:etaZWforR} for the three different choices of $R$.
\begin{table}[t!]
\begin{center}
\renewcommand{\arraystretch}{1.5}
\begin{tabular}{c|cccccc}
$\eta_\ell^{\rm NLO}$ & $R=1$ & $R=0.1$ & $R=0.01$ \\
\hline
$\PZ_{-}\PW^{+}$ & 0.1376(4)  & 0.1361(5) & 0.1338(6) \\ 
\end{tabular}
\caption {
Spin-analysing power $\eta_{\ell}^{\rm NLO}$ extracted from polar-angle distributions for a left-handed $\PZ$ produced in association with an unpolarised $\PW^+$ boson at the LHC (left side of Fig.~\ref{fig:etalNLO_WZ_ZZ}).
Three values of the lepton-dressing radius are considered.
An invariant-mass constraint $81\GeV<M_{\mu^+\mu^-}<101\GeV$ and a collision energy of 13TeV are understood.
}
\label{Tab:etaZWforR}
\end{center}
\end{table}
Similarly to the case of the decay alone, Tab.~\ref{Tab:etacuts}, with a large value of $R$ the extracted value of $\eta_\ell^{\rm NLO}$ gets closer to the value found without cuts, in Eq.~\eqref{eq:etalNLO}, and therefore to 
$\eta_\ell^{\rm eff}$. With a smaller value of $R$,
more photons escape clustering with charged leptons, leading to a slightly larger deviation from the LO picture. Still, the results in Tab.~\ref{Tab:etaZWforR} are different from those in the first row of  Tab.~\ref{Tab:etacuts}. 
 Indeed, as expected, the extraction of the spin-analysing power at NLO EW depends on $R$ but also on the kinematics of the $\PZ$ boson; if the clustering is performed in the laboratory frame, $R$ is not Lorentz-invariant.
If a $\PZ$ boson is produced with a small boost at the LHC, the extracted $\eta_\ell^{\rm NLO}$ will be rather close to the value evaluated as in in Sec.~\ref{sec:decay} with the same value of $R$. On the contrary, in a very boosted kinematic configuration the clustering of photon radiation with charged leptons will be different. This makes the extraction of $\eta_\ell^{\rm NLO}$ inherently process dependent. 

In the right panel of Fig.~\ref{fig:etalNLO_WZ_ZZ} we consider the polar decay angle associated to a left-handed $\PZ$ boson decaying into leptons produced either with an unpolarised $\PW^+$ boson (red curve, same as in the left plot) or with an unpolarised $\PZ$ boson (brown).
The same photon clustering (cone dressing, with $R=0.1$) is used for the two cases. 
Besides the different absolute size of the NLO EW corrections in the two processes, the quadratic fit of the distributions gives rather different predictions for $\eta^{\rm NLO}_{\ell}$, shown in Tab.~\ref{Tab:etaZWandZZ}.
\begin{table}[h!]
\begin{center}
\renewcommand{\arraystretch}{1.5}
\begin{tabular}{c|cccccc}
$\eta_\ell^{\rm NLO}$ & $\PZ_{-}\PW^{+}$ & $\PZ_-\PZ$  \\
\hline
 $R=0.1$ & 0.1362(8)  & 0.1326(7) 
\end{tabular}
\caption {
Spin-analysing power $\eta_{\ell}^{\rm NLO}$ extracted from polar-angle distributions for a left-handed $\PZ$ produced in association with an unpolarised $\PW^+$ or $\PZ$ boson at the LHC (right side of Fig.~\ref{fig:etalNLO_WZ_ZZ}).
An invariant-mass constraint $81\GeV<M_{\ell^+\ell^-}<101\GeV$ and a collision energy of 13 TeV are understood.
\label{Tab:etaZWandZZ}}
\end{center}
\end{table}
 The $\PZ$ boson is slightly more boosted in $\PZ\PZ$ production ($\pt{\PZ} \approx 40\GeV$) than in $\PZ\PW$ ($\pt{\PZ} \approx 25\GeV$), which explains why the $\eta_\ell$ value extracted in the $\PZ\PW$ case ($0.136$) is closer to the value for a $\PZ$ at rest w.r.t.~the one obtained for $\PZ\PZ$ ($0.132$). Another aspect which is worth recalling is that individual helicity states are defined in a specific Lorentz frame \cite{Grossi:2024jae}, therefore the evaluation of the spin-analysing power at NLO EW mildly depends on the chosen coordinate system: a left-handed $\PZ$ boson with the helicity quantised in the laboratory frame and a left-handed $\PZ$ boson with the helicity quantised in another Lorentz frame are produced with different kinematics, so, strictly speaking, the two of them come from different production mechanisms. 

 The overall result of this section is that the correct evaluation of the spin-analysing power $\eta_\ell$ in the presence of higher-order EW corrections has to be carried out case by case. In other words, the presence of real-photon radiation leads to a dependence on the resolution radius of the dressing algorithm and on the mechanism producing the $\PZ$ boson when a cut as the one in Eq.~\eqref{eq:Mmmcut} is present. 
Furthermore, although we have presented results on the dependence on the resolution radius using a standard cone dressing algorithm, we have checked 
 that changing the clustering procedure, \eg the generalised $k_t$ one \cite{Cacciari:2008gp}, leads to small changes in the evaluation of $\eta_\ell^{\rm NLO}$. That said,  we understand that evaluating the corrected $\eta_\ell$ depending on the process and the boost of the $\PZ$ boson in unpractical in an analysis on real data that may impose various phase-space selections. Nevertheless, it is important to estimate such effects and consider them as a source of systematic uncertainty.

\section{Inclusive Z-boson pair production at the LHC}
\label{sec:ZZ}
We can now compare the results obtained via QT and the HA method. We start considering the inclusive production of $\PZ\PZ$ pairs at 13 TeV, with one $\PZ$ boson decaying into muons and the other into an electron-positron pair, as in Eq.~\eqref{eq:WZZZ}. We have performed all calculations that are discussed in this section with the help of the {\sc MoCaNLO} code \cite{Denner:2021csi,Grossi:2024jae,Carrivale:2025mjy}, employing the same input parameters of Sec.~\ref{sec:decay}.  The {\sc MoCaNLO} code enables the direct computation of the spin-density-matrix entries  by selecting specific helicity states for intermediate bosons in Born, real and virtual amplitudes, owing to a flexible interface to the \recola1 library \cite{Actis:2016mpe}. Also, via this code it is possible to perform the calculation either taking into account full off-shell effects or in DPA. Thus, QT  can be applied in both cases and validated against the HA method in DPA.

We focus on the coefficient $\gamma_{1010}$, which is related to the left- and right-helicity content of the diboson process \cite{Grossi:2024jae}. 
Via the HA method it can be written as
 \beq
  \gamma_{1010} =\,-\,\frac{3\,{\eta_\ell}^2}{16\pi}\,\left(f_{--}+f_{++}-f_{-+}-f_{+-}\right)={\eta_\ell}^2 \Deltaf\,,
\label{eq:polar1010}
\eeq
with 
\begin{eqnarray}
 f_{ab} \equiv \rho_{aabb}\,,{~~~\rm and~~~}  \Deltaf\equiv-\,\frac{3}{16\pi}\,\left(f_{--}+f_{++}-f_{-+}-f_{+-}\right)\, .
\end{eqnarray}
As it is manifest, $f_{ab}$ corresponds to the diagonal entry $\rho_{aabb}$ of the spin-density matrix, namely  the joint helicity fraction for two intermediate $\PZ$ bosons with helicities $a$ and $b$, respectively.

We consider massless leptons and apply to both the same-flavour lepton pairs the invariant-mass constraint
\beq
81\GeV < M_{\ell^+\ell^-} < 101\GeV\,,\qquad \ell=\Pe,\mu\,,  \label{eq:Mllcut}
\eeq
as in Eq.~\eqref{eq:Mmmcut}.
Note that in the generic case of two different-flavour lepton-antilepton pairs the factor $\eta_\ell^2$ is replaced by the product ${\eta^{(1)}_\ell}\cdot {\eta^{(2)}_\ell}$, where the two spin-analysing powers take slightly different values owing to possibly different cuts applied to the two $\PZ$-boson virtualities. 
The same consideration would hold if the two fermion-antifermion pairs are characterised by different $SU(2)_L$ and $U(1)_Y$ quantum numbers, \eg in the decay channel with two leptons and two quarks.

The direct calculation at LO and at NLO EW, via the HA method, of the spin-analysing power $\eta_\ell$ and of the quantity $\Deltaf$ (defined in Eq.~\eqref{eq:polar1010}) returns,
\begin{align}
\Deltaf^{\rm LO} = -0.0389(1),\quad
\Deltaf^{\rm NLO} = -0.0388(1),\qquad
\eta^{\rm LO}_\ell = 0.2136(3),\quad
\eta^{\rm NLO}_\ell = 0.1313(7) \,.      \label{eq:DeltaandetaNLO}
\end{align}
Using the same convention adopted in the previous sections for denoting results at LO, NLO EW accuracy and their difference, at NLO EW we obtain 
\begin{eqnarray}
  (\eta^{\rm NLO}_\ell)^2 
 (\Deltaf^{\rm NLO})&=&(\eta^{\rm LO}_\ell+\delta \eta_\ell^{\rm NLO})^2 
 (\Deltaf^{\rm LO}+\delta\Deltaf^{\rm NLO}) \nonumber\\
\label{eq:NLOgamma1010}
 &=&
 (\eta^{\rm LO}_\ell)^2 
 (\Deltaf^{\rm LO}) + 
 2\,\eta^{\rm LO}_\ell\cdot\delta \eta_\ell^{\rm NLO } \cdot \Deltaf^{\rm LO }
 +
 ({\eta^{\rm LO}_\ell})^2 \cdot\delta\Deltaf ^{\rm NLO }+ \mc O(\alpha^2_{\rm ew})\,\\
 &=&\,\gamma^{\rm NLO}_{1010} + \mc O(\alpha^2_{\rm ew})\,. \nonumber
\end{eqnarray}
Thus we can write
\begin{align}
\gamma^{\rm LO}_{1010} = (\eta^{\rm LO}_\ell)^2 
 (\Deltaf^{\rm LO})\,,\qquad
\gamma^{\rm NLO}_{1010} =  \gamma^{\rm LO}_{1010}+ 
 2\,\eta^{\rm LO}_\ell\cdot\delta \eta_\ell^{\rm NLO } \cdot \Deltaf^{\rm LO }
+ ({\eta^{\rm LO}_\ell})^2 \cdot\delta\Deltaf ^{\rm NLO } \,, \label{eq:gammaLOandNLOan}
\end{align}
which lead to the following predictions,
\begin{align}
\label{eq:NLO_LOHA}
{\rm HA~(DPA)}:\qquad\gamma^{\rm LO}_{1010} =  0.001775(1),\qquad
\gamma^{\rm NLO}_{1010} =  0.000407(3)\,.
\end{align}

Applying the QT approach directly on the four-lepton momenta, using the calculation in the DPA, we obtain 
\begin{align}\label{eq:NLO_LOtom}
{\rm QT~(DPA)}:\qquad \gamma^{\rm LO}_{1010}    =  0.001774(5),\quad
\gamma^{\rm NLO}_{1010}   =  0.000404(6)\,.
\end{align}
The values in Eq.~\eqref{eq:NLO_LOHA} and Eq.~\eqref{eq:NLO_LOtom} are in perfect agreement, validating our procedure.

We have also checked that taking into account the full off-shell corrections, including non-resonant effects, and performing the QT,
\begin{align}
{\rm QT~(full)}:\qquad\gamma^{\rm LO}_{1010}    =  0.001708(2),\quad
\gamma^{\rm NLO }_{1010}  =  0.000411(9)\,.    
\end{align}
 the differences with Eq.~\eqref{eq:NLO_LOtom} are very mild.\footnote{We reckon that the good agreement also between the QT (full) and HA (DPA) results is favoured by the presence of the cuts in Eq.~\eqref{eq:Mllcut}, which select the  \emph{on-shell} region. }
 
These  results further confirm that the largest effects for the coefficients $\gamma_{1m1m'}$ originate from the NLO corrections to the spin-analysing power $\eta_\ell$, while minor effects are found for  the spin-density matrix at the production-level. It is manifest from Eq.~\eqref{eq:DeltaandetaNLO} that the quantity $\delta\Deltaf^{\rm NLO}$ is completely negligible. Also, by comparing  Eqs.~\eqref{eq:gammatoCNLO} and \eqref{eq:gammaLOandNLOan}, it is clear that the quantity $\Deltaf$ and $C_{1010}$ are the same quantity with a different normalisation
\beq
C_{1010}=8\pi \Deltaf\,, \label{eq:CandFisthesameperson!}
\eeq
so also NLO EW corrections to  $C_{1010}$ are negligible: $C_{1010}^{\rm NLO}\simeq C_{1010}^{\rm LO}$. 

\medskip

We can now proceed in the opposite direction and test whethere starting from the calculation of $\gamma_{1010}$  with the QT at NLO EW, the correct $C_{1010}$ is obtained. We perform this test using only the DPA results.
First of all, we identify the true value that has be used as a benchmark. Via the HA method we know that 
\beq
C_{1010}^{\rm NLO}\simeq C_{1010}^{\rm LO}=\frac{8 \pi \gamma^{\rm LO}_{1010}}{  (\eta^{\rm LO}_\ell)^2}= 0.977\,.
\eeq

Then, we write different definitions of $C_{1010}$, adopting different choices in the extraction from $\gamma^{\rm NLO}_{1010}$: {\it (1)}
dividing by the LO value of the spin-analysing power $(\eta^{\rm LO}_\ell)^2$,  {\it (2)}  dividing by the NLO value $(\eta^{\rm NLO}_\ell)^2$  or {\it (3)} taking into account the expansion  $(\eta^{\rm NLO}_\ell)^2+2\,\eta^{\rm LO}_\ell\cdot\delta \eta_\ell^{\rm NLO }$. Also, for the last two choices, we look at the difference between {\it (a)} using  $\eta^{\rm NLO}_\ell$ for the process and cuts considered, and the specific recombination-radius $R$ that has been chosen and {\it (b)} the value at the simple decay level without cuts, which corresponds to $\eta^{\rm eff}_\ell$ in Eq.~\eqref{eq:etaleff}:
\begin{eqnarray}
\frac{8 \pi \gamma^{\rm NLO}_{1010}}{  (\eta^{\rm LO}_\ell)^2}&=&0.224\, \label{eq:checkQT1}\qquad{\it (1)}\\
\frac{8 \pi \gamma^{\rm NLO}_{1010}}{  (\eta^{\rm NLO}_\ell)^2}&=&0.593\, \label{eq:checkQT2}\qquad{\it (2a)}\\
\frac{8 \pi \gamma^{\rm NLO}_{1010}}{  (\eta^{\rm eff}_\ell)^2}&=&0.518\,\label{eq:checkQT3}\qquad{\it (2b)}\\
\frac{8 \pi \gamma^{\rm NLO}_{1010}}{  (\eta^{\rm LO}_\ell)^2+ 
 2\,\eta^{\rm LO}_\ell\cdot\delta \eta_\ell^{\rm NLO } }&=&0.977\,\label{eq:checkQT4}\qquad{\it (3a)}\\
\frac{8 \pi \gamma^{\rm NLO}_{1010}}{ (\eta^{\rm LO}_\ell)^2+ 
 2\,\eta^{\rm LO}_\ell\cdot\delta \eta_\ell^{\rm eff } }&=&0.710\, \label{eq:checkQT5}\qquad{\it (3b)}
\end{eqnarray}

The result in Eq.~\eqref{eq:checkQT1} is the standard QT procedure applied to a NLO EW calculation, where using a LO value for the spin-analysing power the NLO EW corrections to this quantity are erroneously propagated to the spin-density matrix $\rho$. The result in Eq.~\eqref{eq:checkQT3} is the procedure proposed in \citere{Goncalves:2025qem}. It is interesting to note that only the procedure in Eq.~\eqref{eq:checkQT4} returns the correct value, which we know since we have calculated directly via the HA method and we know it is equal to the LO value up to per mille effects.

We want to stress that the conclusion of this check is {\it not} that the procedure used in Eq.~\eqref{eq:checkQT4} is the correct one to be applied to data when performing QT. Indeed, data is not at NLO EW accuracy, as it naturally takes into account higher-order effects. This test is precisely saying that the higher-order effects induced by the combined one-loop corrections to both $\PZ$ decays, which effectively consist of two-loop corrections from NNLO EW corrections, are {\it not} negligible. The difference between Eqs.~\eqref{eq:checkQT2} and \eqref{eq:checkQT4} or Eqs.~\eqref{eq:checkQT3} and \eqref{eq:checkQT5} are those $\mathcal{O}(\alpha_{\rm ew}^2)$ effects that have been discarded in the second line of Eq.~\eqref{eq:NLOgamma1010}. In fact, when QT is applied to data, one should use the definition in Eqs.~\eqref{eq:checkQT2}, where $\eta^{\rm NLO}_\ell$ has been calculated for the specific process and cuts considered.  The usage of $(\eta^{\rm eff}_\ell)^2$ is definitely an improvement w.r.t.~$(\eta^{\rm LO}_\ell)^2$, but non optimal. Moreover, as already mentioned before, if different cuts are present for the two $\PZ$ bosons, the quantity $\eta^{\rm NLO}_\mu \eta^{\rm NLO}_e $ should be  used in the normalisation, with $\eta^{\rm NLO}_\ell$ calculated for the $\mu^+\mu^-(e^+e^-)$ pair when $\ell=\mu(e)$.

In the next section we further support the statements of the previous paragraph considering a simplified scenario and we discuss the limitations for the case of the SM Higgs decaying into four leptons. 

\section{Z-boson pairs from a (heavy) Higgs-boson decay}\label{sec:HZZ}
\label{sec:HZZ}

In this section, we consider a system whose spin structure is much more constrained—and therefore simpler—than that of inclusive $\PZ\PZ$ production at the LHC: the decay of a scalar particle into a pair of $\PZ$ bosons. 
Our primary interest is the case where the scalar corresponds to the SM Higgs boson, but we will focus the discussion on the academic scenario in which $\MH > 2 \MZ$, as in the case of possible other scalar particles that are present in new-physics extensions of the SM. The purpose of this study is twofold.

First, in Sec.~\ref{sec:MHoff-shell}, we aim to show that in the case of a heavy Higgs boson with $\MH \gg 2 \MZ$ the procedure outlined in the previous section can be applied to extract the correct spin-density matrix at NLO EW accuracy, since both  $\PZ$ bosons in the $\PH\to \PZ\PZ$ decay are predominantly on-shell. On the contrary, decreasing the value of $\MH$,  off-shell contributions become more and more relevant, reducing the reliability of this approach and rendering it entirely invalid for  $\MH \leq  2 \MZ$, as in the SM.

Second, for the regime  $\MH \gg 2 \MZ$, where both  $\PZ$ bosons are mostly on-shell as in $\PZ\PZ$ production, we perform an approximate NNLO EW calculation by taking into account NLO EW effects in both decays. This test, which is presented in Sec.~\ref{sec:pseudodata}, provides an explicit example that supports the strategy proposed at the end of the previous section for extracting the correct spin-density matrix from data.

In all these sections the calculations for the HA method are performed in NWA,\footnote{For technical limitations the \mocanlo MC framework cannot deal with Higgs-boson decays in the DPA, therefore a simple standalone code based on \recola1.4.4 \cite{Actis:2016mpe} has been devised for the $\PH\rightarrow\PZ\PZ\rightarrow 4\ell$ decay process in the NWA at LO and NLO EW accuracy.} and in general, both for QT and HA results,  we use the same input parameters employed in the previous sections. 

Since the interplay between selection cuts and lepton-photon recombination has been broadly discussed in Sec.~\ref{sec:dress}, the results of this section have been obtained in a simplified setup in that regards. The HA results at NLO properly include (factorisable) virtual and real-photon corrections in the NWA, but the lepton dressing is applied to momenta written in the single-$\PZ$-boson rest frame. In spite of this simplification, the NLO EW accuracy is not spoiled. Furthermore, up to a technical cut ($M_{\ell^+\ell^-}>10\GeV$), we are inclusive on the lepton-pair invariant masses, in order to avoid sensitivity to the details of the dressing algorithm.

\subsection{Off-shell effects in $\PH\to \PZ\PZ$}
\label{sec:MHoff-shell}
If we consider a heavy Higgs boson that shares all the properties of the SM Higgs except for its mass, and examine the decay $H \to \PZ \PZ$, most of the entries of the spin-density matrix of the $\PZ \PZ$ system vanish.
Thus, at variance with generic boson-pair production, the relevant quantum-information observables only depend on few independent $\alpha,\,\gamma$ coefficients. As an example, in the case of the heavy Higgs decay Eqs.~\eqref{eq:polar1010} and \eqref{eq:CandFisthesameperson!} combined together reads
\beq
C_{1010}=8\pi \Deltaf = - 3\,f_{--}\,. \label{eq:CisFisf}
\eeq
Indeed, the structure of the SM scalar-to-gauge coupling leads to the relations $f_{\pm\mp}=0$ and $f_{++}=f_{--}$.
The analogous relation for $C_{111-1}$ is instead 
\begin{align}
C_{111-1}&= - 3\,\Re\,(\mc \rho_{-\rL-\rL})\,.   
\end{align}

\medskip

In the following we consider four representative values for the Higgs-boson mass, 
\begin{equation}
\MH = (183, ~200,~ 225,~ 250) ~{\rm GeV}\,. \label{eq:masses}
\end{equation}
 The condition $\MH>2\MZ$ is valid for all of them, opening up the possibility to produce two on-shell $\PZ$ bosons. QT results at LO and NLO EW accuracy have been obtained with the help of \sloppy {\sc MadGraph5\_aMC@NLO} \cite{Alwall:2014hca,Frederix:2018nkq} simulating the full process 
\beq\label{eq:hto4l}
\PH\rightarrow \Pe^{+}\Pe^{-}\mu^{+}\mu^{-}\,,
\eeq
using the recombination radius $R=1.0$,\footnote{Nevertheless, we have verified that also with $R=0.1$ the results would be indistinguishable with our numerical accuracy. } and requiring $M_{\ell^+\ell^-}>10~\GeV$ for both same-flavour lepton pairs.
Such  an inclusive cut on the lepton-pair invariant masses has been chosen with the purpose of being as close as possible to the HA results obtained  in the NWA,
\beq
\label{eq:htozz}
\PH\rightarrow \PZ(\rightarrow\Pe^{+}\Pe^{-})\,\PZ(\rightarrow\mu^{+}\mu^{-})\,.
\eeq
 
While in the case of inclusive $\PZ\PZ$ production at the LHC the centre-of-mass coordinate system of the four dressed leptons (which we dub ``CM'') is the most natural choice for the spin quantisation, in the case of the Higgs decay the Higgs rest frame (which we dub ``HR'') is often used \cite{DelGratta:2025qyp,Goncalves:2025qem}. The two frames are equivalent at LO, but are different at NLO due to (the small fraction of) unclustered photon radiation. Since the numerical results for the two frames remain compatible at NLO for the observables investigated here, we present only the results in the HR frame.

Following the same procedure discussed in Sec.~\ref{sec:ZZ}, in particular Eqs.~\eqref{eq:NLOgamma1010} and \eqref{eq:gammaLOandNLOan}, we derive the predictions via the HA methods for $\gamma_{1010}$ at LO and NLO accuracy, using for $\eta^{\rm LO}_\ell$ the usual value in Eq.~\eqref{eq:etalLO}, and for
the other inputs the values in Tab.~\ref{Tab:etaandfforMH}.
\begin{table}[t!]
\begin{center}
\renewcommand{\arraystretch}{1.5}
\begin{tabular}{c|cccccc}
$\MH$ [GeV] &$\eta^{\rm NLO}_\ell$ &$f^{\rm LO}_{--}$ & $f^{\rm NLO}_{--}$ \\
\hline
183 &   0.1420(4)& 0.3303& 0.3304  \\
200 &   0.1423(4)& 0.2516&0.2537\\
225 &   0.1432(4)& 0.1619&0.1644 \\
250 &   0.1439(4)&0.1041&0.1059 \\
\end{tabular}
\caption {
Spin-analysing power $\eta_{\ell}^{\rm NLO}$ calculated at NLO EW and helicity fraction $f_{--}$ calculated at LO and NLO EW,  for the different $\MH$ benchmark points.
MC-integration uncertainties are shown in parentheses, if not shown they are smaller than $10^{-4}$.
\label{Tab:etaandfforMH}}
\end{center}
\end{table}
As can be noticed, the value of $\eta^{\rm NLO}_\ell$ depends on the value of $\MH$, which enters via loop corrections. Both the helicity fractions and $\eta^{\rm NLO}_\ell$ are computed at exact NLO EW corrections.

The values of $\gamma_{1010}$ obtained via the HA method for the four benchmark masses in \eqref{eq:masses} are  listed in Table~\ref{tab:HH_Gamma_NLO} and compared with the corresponding results obtained via QT. We note  from the previous equations that also for this process the values of $\Deltaf$, and therefore of also the ``true values'' of $C_{1010}$, receive very mild NLO EW corrections. 

\begin{table}[t!]
\begin{center}
\renewcommand{\arraystretch}{1.5}
\begin{tabular}{C{2cm}|C{2.3cm}C{2.3cm}|C{2.3cm}C{2.3cm}}
\hline
 \multicolumn{5}{c}{$\gamma_{1010}$ coefficient ($\times 10^3$)} \\[0.1cm]
\hline
& \multicolumn{2}{c|}{QT} & \multicolumn{2}{c}{HA} \\[0.1cm]
\hline
$\MH$                       & LO &  NLO &  LO  &   NLO  \\ 
\hline
183 GeV & $-1.5506(9)$ & $-0.44(1)$ & $-1.790(1)$ &  $-0.596(6)$ \\
200 GeV & $-1.3085(5)$ & $-0.417(9)$ & $-1.364(1)$ & $-0.469(5)$  \\
225 GeV & $-0.8601(7)$ & $-0.294(7)$ & $-0.8776(7)$ & $-0.315(3)$  \\
250 GeV&  $-0.559(1)$  & $-0.200(3)$ & $-0.5643(4)$  &  $-0.208(2)$  \\\hline
\end{tabular}
\caption {Numerical results for the $\gamma_{1010}$ coefficient (inflated by a factor $10^3$) extracted via QT from full off-shell simulations and via HA methods by direct calculation with intermediate on-shell $\PZ$ bosons, using Eqs.~\eqref{eq:NLOgamma1010} and \eqref{eq:gammaLOandNLOan} and inputs from Tab.~\ref{Tab:etaandfforMH}.
The $\PZ$ bosons are identified by the flavour of the decay products and the spin-quantisation axis is given by the direction of the $\PZ$ decaying to the $ \Pe^+ \Pe^-$ pair in the Higgs reference frame (HR). 
MC-integration uncertainties are shown in parentheses. }
\label{tab:HH_Gamma_NLO}
\end{center}
\vspace{-5mm}
\end{table}

The $\gamma_{1010}$ coefficient is generally small, as in $\PZ\PZ$ production, and decreases for large Higgs-boson masses. Indeed,   at large $\MH$ the longitudinal polarisations of the two $\PZ$ bosons is favoured, both at LO and NLO EW. Since $f_{\rL\rL}$ increases, $f_{--}$ must decrease.

For masses above $\MH =225\GeV$, although $\PZ$ off-shell effects are taken into account only in the QT result, the HA method, which is performed in NWA,  reproduces very well both the LO and NLO predictions. While a fairly good agreement is found also for the intermediate cases, the Higgs mass $\MH = 183\GeV\gtrsim 2\MZ$ leads to a larger discrepancy. We notice that the discrepancy ($\rm QT/HA=0.87$ at LO) is even larger at NLO ($\rm QT/HA=0.77$). 

\begin{figure}
    \centering
    \includegraphics[width=0.46\textwidth]{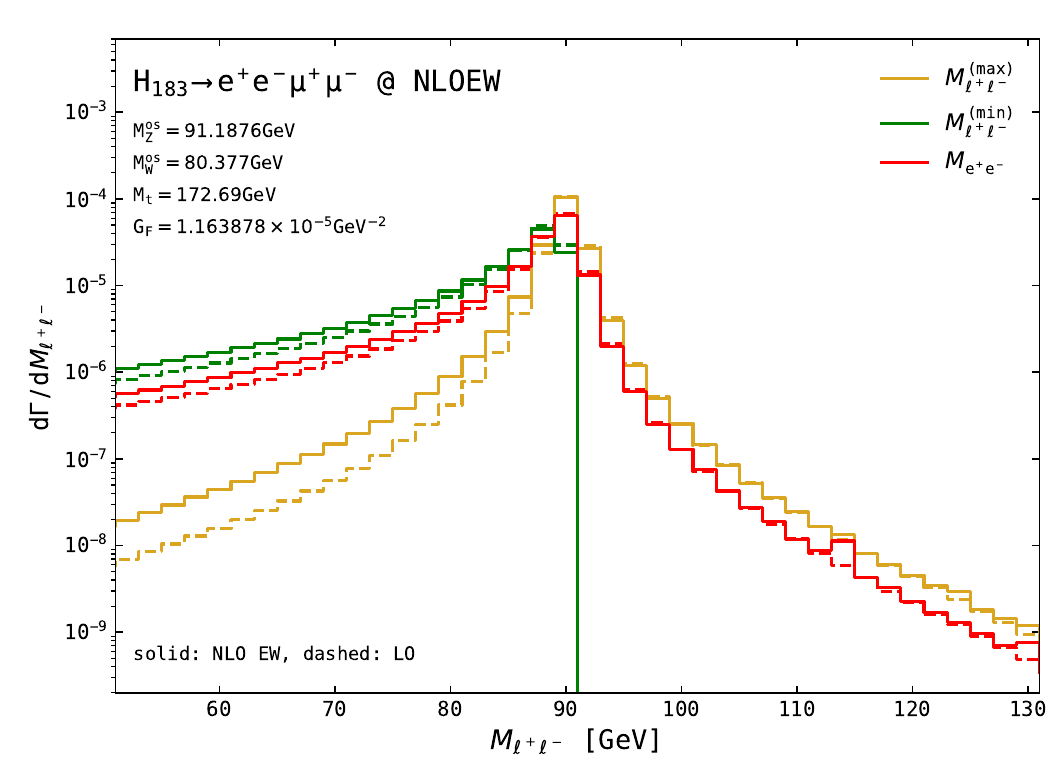}
    \includegraphics[width=0.46\textwidth]{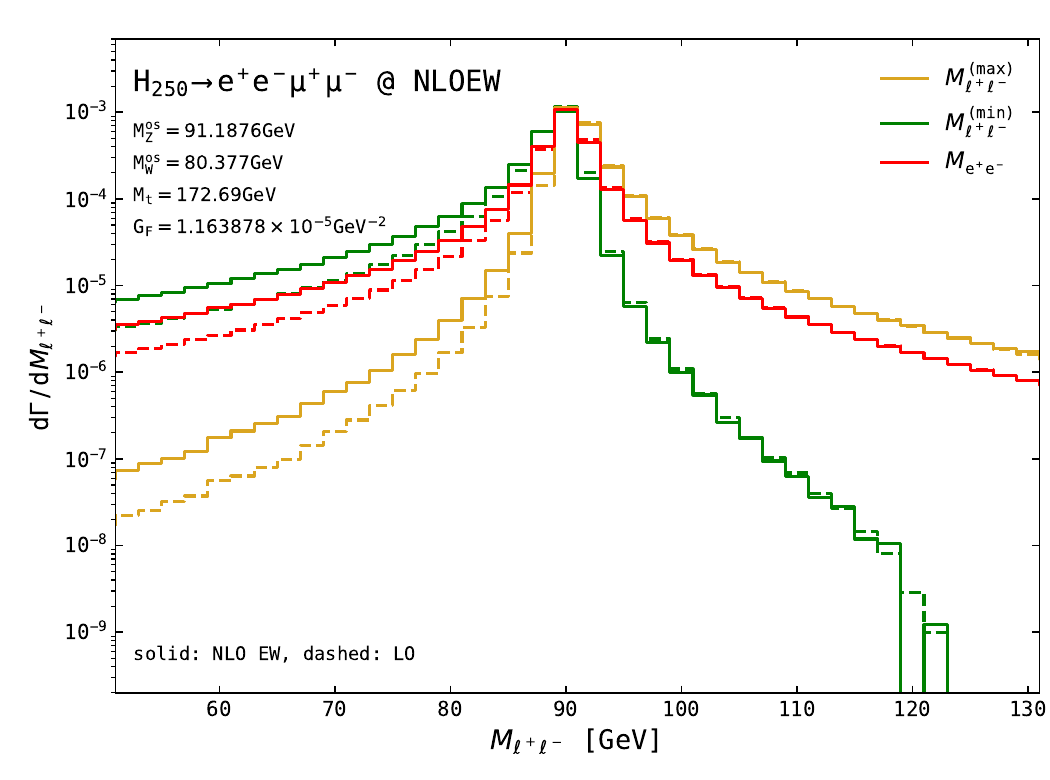}
    \caption{Invariant-mass distributions of the same-flavour lepton pair with the largest/smallest invariant mass (yellow/green) and of the electron-positron pair (red) in the $\PH\rightarrow e^{+}e^{-}\mu^{+}\mu^{-}$ off-shell process, for $\MH = 183\GeV$ (left) and $\MH = 250\GeV$ (right). Solid curves: NLO EW, dashed curves: LO. A lepton-dressing resolution radius of $R=0.1$ is understood at NLO EW.}
    \label{fig:masses}
\end{figure}

When the mass of the Higgs is just above the value $2\MZ$, the two $\PZ$ bosons  emerging from the decay are almost at rest in the Higgs rest frame. Thus, if the two  $\PZ$ bosons are allowed to be off-shell, a sizeable fraction of events will feature at least one of the lepton pairs that is {\it not} on-shell. This feature can be easily seen in the left plot of Fig.~\ref{fig:masses} where  we compare, for $\MH=183~\GeV$,  the invariant-mass spectrum  at NLO (solid lines) and LO (dashed lines) for three different lepton-pair definitions. Red lines refer to  the lepton pair identified via its specific flavour,  which for definiteness we choose as $e$: $M_{e^+e^-}$. Ochre (Green) lines refer to the same-flavour lepton pair with the largest (smallest) invariant mass: $M_{\ell^+\ell^-}^{\rm max}$ ($M_{\ell^+\ell^-}^{\rm min}$).

We notice that the typical Breit-Wigner line-shape   is present only for the distribution of $M_{\ell^+\ell^-}^{\rm max}$. Instead, in the case of $M_{e^+e^-}$ many  events are off-shell, with lower values than $\MZ$, and even more for $M_{\ell^+\ell^-}^{\rm min}$. In other words, for $M_{e^+e^-}$ and $M_{\mu^+\mu^-}$, which are identical, the off-shell fraction is very large and even larger at NLO w.r.t.~the LO case. On the contrary, the off-shell fraction for $M_{\ell^+\ell^-}^{\rm max}$ is very small, while $M_{\ell^+\ell^-}^{\rm min}$ is almost always off-shell.

Looking at the right plot of Fig.~\ref{fig:masses}, we find  the same information as in the left one but for $\MH=250~\GeV$, showing a very different picture. The $M_{e^+e^-}$ distribution features the typical Breit-Wigner lines-shape, while the $M_{\ell^+\ell^-}^{\rm max}$ ($M_{\ell^+\ell^-}^{\rm min}$) distribution is tilted towards large (small) invariant-mass values. Thus, a large fraction of events have both $M_{\ell^+\ell^-}^{\rm max}$ and $M_{\ell^+\ell^-}^{\rm min}$ pairs that are off-shell. On the contrary, both $M_{e^+e^-}$ and $M_{\mu^+\mu^-}$ are typically on-shell. 

The QT results for $\MH=183~\GeV$, unlike those for high masses, are biased by the presence of non-resonant effects, which therefore affect the  two-qutrit interpretation of the angular coefficients. One could significantly mitigate this effect  by applying strict cuts around $\MZ$ on the same-flavour lepton pairs, but this condition would not be anyway possible for the case $\MH=125~\GeV$.
\footnote{The coordinate system is defined with the spin-quantisation axis aligned to the $\PZ$ boson labeled $V_1$. Since each $\PZ$ is reconstructed from same-flavor leptons, the only ambiguity is  in the assignment of $V_1$. Here, we follow \citere{Grossi:2024jae} and assign $V_1$ according to the flavor of its decay leptons. Alternatively, $V_1$ can be chosen as the lepton pair with the largest invariant mass \cite{DelGratta:2025qyp,Goncalves:2025qem,Aguilar-Saavedra:2025byk}. As shown in Fig.~\ref{fig:masses}, this choice induces an asymmetry between $V_1$ and $V_2$ ($\alpha^{(1)}_{lm}\neq \alpha^{(2)}_{lm}$), complicating interpretation without invariant-mass constraints. For $\MH=183~\GeV$, both definitions have drawbacks: the flavor-based choice yields asymmetric invariant-mass distributions for $V_1$ and $V_2$, while the mass-based choice enhances their difference.}

\medskip

In conclusion, while it is tempting to apply the proposed strategy outlined at the end of Sec.~\ref{sec:ZZ} to the decay of a SM Higgs boson to four charged leptons, this is not possible owing to the unavoidable large off-shell effects. The SM Higgs mass forbids the presence of two intermediate on-shell bosons and therefore critically affects the spin interpretation of the results extracted through QT \cite{Grossi:2024jae,DelGratta:2025qyp}. At $\MH=250~\GeV$, the approach would work, but already with $\MH=183~\GeV$, where both $\PZ$ bosons can still be in principle on-shell, this strategy starts to degrade. It is therefore natural to expect that this strategy would completely fail for $\MH=125~\GeV$. In addition, no sensible way to calculate  the ``true'' $C_{1010}$ value is possible. 
Nonetheless, the production of two qutrits in a spin-singlet state, {\it i.e.}~from the decay of a scalar, is very interesting in light of a quantum-information interpretation. Several studies have indeed shown how this final state could be used to study entanglement, Bell non-locality, and multipartite entanglement~\cite{Barr:2021zcp,Aguilar-Saavedra:2022wam,Ashby-Pickering:2022umy,Aguilar-Saavedra:2022mpg,Fabbrichesi:2023cev,Fabbrichesi:2023jep,Aoude:2023hxv,Bernal:2023ruk,Fabbri:2023ncz,Morales:2023gow,Bernal:2024xhm,Grossi:2024jae,Sullivan:2024wzl,Wu:2024ovc,Grabarczyk:2024wnk,Aguilar-Saavedra:2024jkj,Subba:2024aut,DelGratta:2025qyp,Ding:2025mzj,Aguilar-Saavedra:2025byk,Goncalves:2025qem,Ruzi:2025jql,Goncalves:2025xer}. Given our findings, we further confirm the message of \citere{DelGratta:2025qyp}: modifications of the definition of quantum-information observables have to be employed, so that they do not depend on the $C_{1m1-m}$ coefficients.

\medskip

\begin{table}[h!]
\begin{center}
\renewcommand{\arraystretch}{1.5}
\begin{tabular}{C{3.7cm}||C{2.3cm}C{3.7cm}C{3.7cm}}
QT ($\MH=250\GeV$) & LO & NLO from Eq.~\eqref{eq:checkQT1} & NLO from Eq.~\eqref{eq:checkQT4} \\
\hline\hline
$C_{1010}$ & $-$0.3094(7) &  $-$0.114(4)  & $-$0.315(7)\\ 
$C_{111-1}$ & 0.849(1) & 0.319(4) &   0.92(1) \\ \hline
 & LO & NLO  &  
\\ \hline 
$C_{222-2}$ & 0.3106(2) & 0.3130(4) &  -  \\ 
$C_{212-1}$ & $-$0.8508(1) & $-$0.8531(3)  & -  \\
$C_{2020}$ & 1.6895(2) & 1.6682(6) & - \\  \hline
\end{tabular}
\caption {
Spin-correlation coefficients $C_{lml'm'}$ extracted with QT from $\PH\rightarrow \Pe^+\Pe^-\mu^+\mu^-$ events at LO and at NLO EW accuracy. A Higgs-boson with $\MH=250\GeV$ is considered. MC-integration uncertainties are shown in parentheses.}
\label{tab:HH_C_NLO_QTnew}
\end{center}
\vspace{-5mm}
\end{table}

The correct value of $\eta_\ell$ is mandatory for a reliable extraction of $C$ coefficients not only for $\PZ\PZ$ production, but whenever the $\PZ$ bosons are on-shell. We explicitly show it in  another case:  an hypothetical beyond-the-SM heavy scalar. It is important to keep in mind that besides their relevance in the context of quantum-information, it has been shown the spin-density matrix coefficients  can also leverage the sensitivity in the searches for new physics \cite{Fabbrichesi:2023jep,Bernal:2023ruk,Aoude:2023hxv,Sullivan:2024wzl,Subba:2024aut,DelGratta:2025qyp}.

In Tab.~\ref{tab:HH_C_NLO_QTnew} we show the $C_{1010}$ and $C_{111-1}$ coefficients extracted from a LO simulation via QT and also from an NLO simulation, either following the same procedure at LO,  namely Eq.~\eqref{eq:checkQT1}, or considering the correct value of $\eta_\ell$ and taking into account the expansion in powers of $\alpha_{\rm ew}$,  namely Eq.~\eqref{eq:checkQT4}. We also show the $C_{222-2}$, $C_{212-1}$ and $C_{2020}$ coefficients, where the $\eta_\ell$ dependence is not present and no ambiguity is present. The large effects (falsely) induced by NLO EW corrections to $C_{1010}$ and $C_{111-1}$, and that are visible in the second column of numerical values, are not present in the third column, similarly to what is already observed in $C_{222-2}$, $C_{212-1}$ and $C_{2020}$ coefficients. 

Using a proper value of $\eta_\ell$ is paramount for a reliable extraction of the $C$ parameters. We stress that the conclusion  is {\it not} that the procedure used in Eq.~\eqref{eq:checkQT4} is the correct one to be applied on data when performing QT. We further explore this fact in the next section.

\subsection{Optimal approach for extracting $C$ coefficients from pseudo-data}
\label{sec:pseudodata}

In Sec.~\ref{sec:ZZ} we have shown that performing an NLO EW calculation, the correct value of the $C_{1m1m'}$ coefficients can be extracted via QT only using the NLO EW value of $\eta_\ell$, and the prescription in Eq.~\eqref{eq:checkQT4}. An analogous result has been demonstrated for the case of the Higgs decay with $\MH=250 \GeV$, in Tabs.~\ref{tab:HH_Gamma_NLO} and \ref{tab:HH_C_NLO_QTnew}. We argued in the discussion around Eqs.~\eqref{eq:checkQT1}--\eqref{eq:checkQT5} that using the procedure in Eq.~\eqref{eq:checkQT4} rather than Eq.~\eqref{eq:checkQT2} is a signal that higher-order effects are important, and that  the latter should be used with real data. In the following we further support these statements, considering the case of an Higgs decay with  $\MH=250 \GeV$.

The first step in our argument consists in identifying a procedure in order to take into account higher-order effects and therefore simulate as close as possible, for what concerns higher-order corrections,  pseudo-data. From discussions in the previous sections it is clear that while NLO EW corrections to the $\rho$ density matrix are minimal,\footnote{It can be seen by comparing LO and NLO values in Eq.~\eqref{eq:DeltaandetaNLO} for $\Deltaf$ and in Tab.~\ref{Tab:etaandfforMH} for $f_{--}$.} the NLO EW corrections to the $\PZ$-boson decay into leptons, and especially $\eta_\ell$, are anomalously large. Thus, it is important to take into account NLO EW corrections to the decay of both $\PZ$ bosons in a multiplicative way, {\it i.e.}, as the r.h.s.~of Eq.~\eqref{eq:RhoGammaGammaLO} but with all quantities at NLO EW accuracy.

In order to achieve the aforementioned accuracy, we have devised a simplified approach that captures the dominant effects. We illustrate it by simulating the $\PH\rightarrow \Pe^{+}\Pe^{-}\mu^{+}\mu^{-}$ process at LO and for $\MH=250\GeV$, using the $\{\GF,\,\MZ,\,\sweff\}$ input EW scheme \cite{Kennedy:1988sn,Renard:1994ay,Ferroglia:2001cr,Ferroglia:2002rg,Chiesa:2019nqb,Amoroso:2023uux,Biekotter:2023vbh,Chiesa:2024qzd}, with $\GF$ and $\MZ$ as in \refse{sec:decay} and 
\begin{equation}\label{eq:sthetaeff}
\sweff=0.23192\,. 
\end{equation}
Note that the input value of the mixing angle in Eq.~\eqref{eq:sthetaeff} has been chosen equal to the value obtained in the $G_\mu$ scheme including one-loop EW corrections (for $\MH=250\GeV$) as done in \refse{sec:decay}. In this way, the simulation effectively takes into account NLO corrections to both decays at the same time, without discarding the $\mc O(\alpha^2_{\rm ew})$ terms of Eq.~\eqref{eq:NLOgamma1010}, unlike what has been done so far following Eq.~\eqref{eq:gammaLOandNLOan}. The QT analysis of the results of this simulation can be compared with a value for the HA method, which can be derived in the $G_\mu$ scheme employing directly the l.h.s.~of Eq.~\eqref{eq:NLOgamma1010}, without discarding $\mc O(\alpha^2_{\rm ew})$ terms.

We dub the aforementioned approximation aNNLO, as ``approximate NNLO'', and we find the following results:
\begin{eqnarray}
{\rm QT:}\qquad \gamma^{\rm aNNLO}_{1010}  & =&  -0.000262(6)\,,\\
{\rm HA:}\qquad\gamma^{\rm aNNLO}_{1010}    &= &-\frac3{8\pi}\left(\eta_\ell^{\rm NLO}\right)^2 \cdot f_{--}^{\rm NLO}= -0.000262\,. \label{eq:HANNLO}
\end{eqnarray}
We find a very good agreement between the two procedures and therefore we can derive the following statements:

\begin{itemize}
\item{From Tab.~\ref{Tab:etaandfforMH} we see that at $\MH=250~\GeV$ the LO prediction for $\gamma_{1010}$ receives about $ -63\%$ corrections at NLO EW. If the comparison is instead done between the LO and aNNLO predictions, corrections amount at $ -54\%$, {\it i.e.}, the approximated aNNLO EW corrections are roughly $ +10\%$ of the LO prediction and especially $ +26\%$ of the NLO EW prediction. Such effects must be taken into account, not only for the sake of precision.  }
\item{Since $f_{--}$ is directly related to $C_{1010}$, as shown in Eq.~\eqref{eq:CisFisf}, and we see explicitly in Eq.~\eqref{eq:HANNLO} that it is proportional to $\left(\eta_\ell^{\rm NLO}\right)^2$, in order to extract the correct value of $C_{1010}$ the procedure to follow is the one in Eq.~\eqref{eq:checkQT2}, {\it i.e.}, precisely dividing by $\left(\eta_\ell^{\rm NLO}\right)^2$. We would obtain the value $C^{\rm aNNLO}_{1010}=-0.309(8)$, which by construction is in agreement with the LO and NLO values in Tab.~\ref{tab:HH_C_NLO_QTnew}.}
\end{itemize} 

Thus, we clearly see that, for processes in which the $\PZ$ bosons are on-shell and decay into leptons, higher-order EW effects can be sizeable and must be taken into account.
 The proper extraction of  the $C_{1m1m'}$ coefficients must be performed following Eq.~\eqref{eq:checkQT2}, possibly taking into account the effects of cuts on the $\PZ$ decay products, as discussed in Sec.~\ref{sec:dress}.

\section{Conclusions and outlook}\label{sec:conclu}

In this work we have carried out a systematic study of the quantum-tomography (QT) extraction of the $\PZ$-boson spin-density matrix at next-to-leading-order (NLO) electroweak (EW) accuracy.  
Our analysis elucidates the origin of the unexpectedly large NLO effects reported in earlier studies \cite{Grossi:2024jae,DelGratta:2025qyp,Goncalves:2025qem} and provides a robust framework for incorporating EW corrections into the determination of spin-correlation observables.

First, we have computed the complete NLO EW corrections to the decay matrix of a single on-shell $\PZ \to \ell^+\ell^-$ transition. We have shown that the analytic structure of the matrix remains  unaltered with respect to the leading-order (LO) expression, while the spin-analysing power $\eta_\ell$ receives a sizeable negative shift of approximately $-35\%$. This effect, mostly coming from loop corrections, accounts for the dominant NLO impact observed in previous QT-based analyses. 

Second, we have investigated the implications of photon radiation and lepton-dressing algorithms, particularly in boosted kinematics relevant for LHC processes, in conjunction with cuts on the same-flavour lepton-pair masses. We demonstrated that the extraction of $\eta_\ell$ depends not only on the dressing radius but also on the production mechanism and the kinematic configuration of the $\PZ$ boson. This highlights the need for a process- and setup-specific determination of $\eta_\ell$ when applying QT in realistic collider environments.

Third, we applied our refined prescription to inclusive $\PZ\PZ$ production at the LHC and showed that QT and the helicity-amplitude (HA) approach yield results in excellent agreement when consistently using the NLO-corrected value of $\eta_\ell$. Full off-shell effects, while present, were found to have only a mild numerical impact under typical fiducial selections.

Finally, we studied Higgs-boson decays to four leptons. For a Standard Model Higgs with $\MH \simeq 125\,\text{GeV}$, the off-shell effects in the $\PH \to \PZ\PZ^* \to 4\ell$ channel prevent the straightforward application of the same prescription. Conversely, for a heavy Higgs boson ($\MH \gg 2 \MZ$), the approach remains valid  and we quantified the impact of formally higher-order (NNLO EW) effects, confirming their relevance in precision studies. This implies that  order $\mc O(\alpha^2_{\rm ew})$ effects have to be taken into account  when  spin-density-matrix coefficients are extracted from data. 

Our results establish a robust and general framework for performing quantum tomography at NLO EW accuracy for any process involving one or more {\it on-shell} $\PZ$ bosons.  EW higher-order effects can be sizeable and have to be taken into account: the proper extraction of spin-correlation coefficients that involves the spin-analysing power $\eta_\ell$ must be performed using its NLO EW prediction, possibly taking into account, via the HA method, the effects of cuts on the $\PZ$ decay products.

We stress that the relevance of our findings does not pertain uniquely the measurement of quantum-entanglement and Bell-non-locality markers, but in general the sound determination of the spin-density matrix of any $\PZ+X$ system, with the $\PZ$ boson decaying into charged leptons. The results of this work provide clear guidance for upcoming experimental analyses both at the LHC and at future $\Pe^+\Pe^-$ facilities, where the clean experimental environment and high luminosity will allow measurements with unprecedented accuracy. Looking forward, the extension of this study to mixed QCD--EW corrections, to parton-shower effects, and to multi-boson final states  will be essential to fully exploit quantum-inspired measurements at high-energy colliders, both within and beyond the SM.

\section*{Acknowledgements}
The authors acknowledge support from the European Union (EU) COMETA COST Action (CA22130). 
GP acknowledges financial support from 
the EU Horizon Europe research and innovation programme
under the Marie-Sk\l{}odowska Curie Action (MSCA) ``POEBLITA - POlarised Electroweak Bosons at the LHC with Improved Theoretical Accuracy'' - grant agreement Nr.~101149251 (CUP H45E2300129000) 
and
the Italian Ministry of University and Research (MUR), with
EU funds (NextGenerationEU), through the 
PRIN2022 grant agreement Nr.~20229KEFAM (CUP H53D23000980006). 
FF acknowledges financial support from 
the EU Horizon Europe research and innovation programme
under the MSCA ``QUANTUMLHC - Exploring quantum observables at the LHC'' - grant agreement Nr.~101107121 (CUP J33C23001080006). 
MG is supported by CERN through the CERN Quantum Technology Initiative.
DP acknowledges the financial support by the MUR, with
NextGenerationEU funds, through the PRIN2022
grant Nr.~2022EZ3S3F; likewise FM, through the PRIN2022 grant Nr.~2022RXEZCJ,
and by the project ``QIHEP--Exploring the foundations of quantum information
in particle physics'', which is financed through the PNRR with
NextGenerationEU funds, in the context of the extended
partnership PE00000023 NQSTI (CUP J33C24001210007).

\bibliographystyle{JHEPmod}
\bibliography{polvv}

\end{document}